\documentclass[iop,twocolumn,tighten]{aastex701}
\usepackage{graphicx,multirow,url,color,amsmath}
\usepackage{makecell}
\usepackage{tabularx}
\usepackage[normalem]{ulem}
\usepackage{CJK}
\usepackage{lineno}
\linenumbers

\hypersetup{
  colorlinks,
  linkcolor={blue!88!black!80},
  citecolor={blue!88!black!80},
  urlcolor={blue!88!black!80}}

\newcommand{\msun}{{M_\odot}}
\newcommand{\mbh}{{M_{\rm BH}}}
\newcommand{\mdotbh}{{\dot{M}_{\rm BH}}}
\newcommand{\ergs}{{\rm erg\,s^{-1}}}

\begin{document}
\begin{CJK*}{UTF8}{gbsn} 

\shorttitle{Theoretical Radio Size of LLAGNs}
\title{Radio Core Size of Low-luminosity Active Galactic Nuclei under the MAD-jet model}

\author{Yue Xu (许悦)}
\affiliation{Shanghai Astronomical Observatory, Chinese Academy of Sciences, 80 Nandan Road, Shanghai 200030, People's Republic of China}
\affiliation{School of Astronomy and Space Sciences, University of Chinese Academy of Sciences, 19A Yuquan Road, Beijing 100049, People's Republic of China}
\email[]{xxx@}

\author[0000-0001-9969-2091]{Fu-Guo Xie (谢富国)}
\altaffiliation{fgxie@shao.ac.cn}
\affiliation{Shanghai Astronomical Observatory, Chinese Academy of Sciences, 80 Nandan Road, Shanghai 200030, People's Republic of China}
\email[]{fgxie@shao.ac.cn}

\date{Accepted XXX. Received YYY; in original form ZZZ}

\begin{abstract}
After decades of efforts, there are now fruitful high-resolution radio observations of low-luminosity active galactic nuclei (LLAGNs), and the observed frequency has extended from $\sim$10 GHz up to $\sim$200 GHz. In this work, based on a model that combines a magnetically arrested disk (MAD) and a Blandford-Znajek-like jet, we carried out detailed analysis on size and location of the radio core of LLAGNs. The radio core size of nearby LLAGN M104 is re-visited based on this new model. We successfully reproduce a $size\propto\nu^{-1}$ scaling between 1 GHz and tens of GHz, if more than $50\%$ of electrons in jet follow a power-law (PL) distribution. We further confirm that, at high radio frequencies emission from MAD exceeds that from jet, and a flatter size-frequency slope is observed. The impact of PL electrons in MAD is also investigated. For those $L_{\rm bol}/L_{\rm Edd} \gtrsim (3-8)\times 10^{-6}$ LLAGNs and black hole binaries in their hard state, PL electrons are expected to be highly suppressed due to strong radiative cooling (so-called `synchrotron boiler' effect). 
\end{abstract}




\section{Introduction}\label{sec:intro}

More than 40\% of nearby galaxies host low-luminosity active galactic nuclei (LL-AGNs) at their center \citep{Ho1997}. These LLAGNs exhibit a distinct spectral energy distribution (SED) compared to those bright AGNs (see \citealt{Ho2008} for a review). Significant progress has been made over the past $4-5$ decades on the structure, the dynamics and the radiative properties of AGNs \citep{Antonucci1993, Urry1995}. The accretion system in LLAGNs can be broadly decomposed into three key components (for reviews, see e.g., \citealt{Ho2008,Yuan2014}): a standard thin accretion disk \citep{shakura1973} that is located some distance away from the supermassive black hole (SMBH); a geometrically thick but optically thin hot accretion flow (e.g., \citealt{Narayan1994,Yuan2014}) that is located within this distance, and a relativistic jet launched from close to SMBH and in a direction perpendicular to the plane of the hot accretion flow. Surrounding them are broad-line region, narrow-line region, and dusty torus \citep{Urry1995}. Recent observations of two nearby LLAGNs, e.g., Sgr A* and M87 \citep{EHT2019, EHT2022}, suggest that the hot accretion flow component in LLAGNs may be a magnetically-arrested disk (MAD; \citealt{Narayan2003}), i.e., its magnetic flux can be accumulated and amplified to a saturated level such that the magnetic field strength is comparable to the gravity of SMBH.

Resolving and understanding the structure of accretion disk has now been one of the frontiers in AGN physics. Part of this can be probed indirectly through timing/variability analysis of AGN light curves. A direct and more reliable approach is to carry out high spatial resolution observations. In X-rays, {\it Chandra} has a resolution of 0.5\,arcsec, which only resolves $\sim0.24$\,kpc within 100\,Mpc distance.\footnote{For the local Universe we focus on, the physical resolution $a$ (in pc) can be estimated as, $a\approx24\,{\rm pc}\,(D/10\,{\rm Mpc})\,({\rm Resol.}/0.5'') \approx 4.8\times10^{-2}\,{\rm pc}\,(D/10\,{\rm Mpc})\,({\rm Resol.}/1\,{\rm mas})$, where $D$ is the luminosity distance, and mas is milli-arcsecond.} In optical, the Very Large Telescope Interferometer (VLTI) uses interferometry to combine the light collected by the four VLT telescopes to achieve a resolution of milli-arcsecond (mas). The interferometry has been widely adopted in the radio community, and a resolution as high as mas or $\mu$as can be reached.

Despite continuous efforts over decades, spatially resolving the structure of AGN remains a challenging task. For an example, at centimeter radio band, although the earth-size Very Long Baseline Array (VLBA) and European VLBI Network (EVN) can achieve a resolution of mas, the observed angular size still suffers broadening due to interstellar scattering at low frequencies (e.g., \citealt{Bower2004}). Such broadening becomes negligible at high frequencies, thus intrinsic size can be measured directly. High-resolution and high-sensitivity capabilities at submillimeter band have significantly improved, i.e., the Atacama Large Millimeter/submillimeter Array (ALMA) reaches an angular resolution ranging from 75 mas to 25 mas (e.g., \citealt{ALMA2015}). The Event Horizon Telescope (EHT) that includes ALMA achieves a resolution of 20 $\mu$as at 230 GHz (e.g., \citealt{EHT2019}), and recently the first very long baseline interferometry (VLBI) at 345 GHz achieves an angular resolution of 19 $\mu$as (e.g., \citealt{Raymond2024}). These advances provide the capability of resolving the $\lesssim 100\,R_{\rm g}$ region of nearby SMBH (here $R_{\rm g} = GM_{\rm BH}/c^2$ defines the gravitational radius of an  $M_{\rm BH}$-mass SMBH).

High-resolution observations of LLAGNs not only measure their flux but also reveal their locations and detailed structures. The bright optically-thick radio core of the jet in LLAGNs (e.g.,  \citealt{Sokolovsky2011,Kutkin2014}) shows a shift of position with observed frequency, a phenomenon called ``core shift''. Physically this reveals that the synchrotron self-absorption optical depth $\tau\sim1$ is not a constant but varies with frequency \citep{Rybicki1979}, i.e., the self-absorbed synchrotron emission of jet peaks at a distance $z\propto \nu^{-1}$ along the jet (where $\nu$ is the observed frequency, \citealt{Blandford1979, Konigl1981, Zdziarski2015}). More generally, by a combined analysis of both spectral and (resolved) size properties of LLAGNs, several key problems of BH accretion physics can be studied (e.g., \citealt{BandX2019}), i.e., origin and decomposition of radio emission at different frequencies, jet composition (e.g., hadronic versus leptonic, dominance of non-thermal\footnote{In this work, non-thermal is used equally/interchangeably with (broken) power-law (PL).} electrons/positrons), radiative contribution of non-thermal electrons in hot accretion flow (e.g.,  \citealt{Mahadevan1997, BandX2019, Kimura2020}), etc. Non-thermal electrons are also crucial to understand high-energy neutrinos from blazar TXS 0506+056 \citep{Yang2025}.

In this work, we investigate the radio size, as well as the broadband spectrum, of LLAGNs under a coupled magnetically-arrested disk (MAD, \citealt{Narayan2003}) and jet model (hereafter MAD-jet model), which recently has been applied to the broad spectrum of LLAGN M87 \citep{Kuze2024}. In Section~\ref{sec:model}, we present the MAD-jet model and its basic setup.  In Section \ref{sec:m104case}, we take the well-observed nearby LLAGN M104 as a testbed, where we provide a detailed modelling of both the SED and the radio core size. We then in Section~\ref{sec:size_LLAGN_MAD.th} take values of several unconstrained parameters of M104 as fiducial ones, and apply the MAD-jet model to LLAGNs in general. Here observation of the Fundamental Plane (FP) of black hole activity is adopted to determine the MAD-jet coupling. In Section~\ref{sec:size_LLAGN_MAD.pl}, we investigate the impact of non-thermal electrons in MADs. Finally in Section \ref{sec:summary} we conclude with a brief summary.

\section{model and basic setup}\label{sec:model}

In this work, we focus on a coupled model that includes both a magnetically-arrested disk (MAD, \citealt{Narayan2003}) and a jet that launches from the central BH and in a direction perpendicular to the MAD (or aligned to BH spin, see \citealt{Liska2018} for numerical simulations.). Basic model parameters, together with their meanings, are listed in Table \ref{tab:m104fit}.

\subsection{Magnetically-Arrested Disk (MAD)} \label{sec:mad}
During the early development of hot accretion flow in sub-Eddington systems \citep{Narayan1994}, the net magnetic flux supplied by accreting gas is assumed to be weak. Consequently, the magnetic flux near BH is moderately low. It was not until the 2000s that the importance of the supply and the accumulation of magnetic flux in accretion disks was recognized, and the system with saturated magnetic fields is called a magnetically-arrested disk (MAD, \citealt{Narayan2003}). MAD is theoretically plausible, as a large coherent magnetic flux in the interstellar medium (feeding source for accretion) near central AGN is observed (e.g., \citealt{Zamaninasab2014}). Transverse Faraday rotation gradients in jets, also provides direct evidence of global/ordered magnetic fields near SMBH (e.g., \citealt{Lisakov2021, Contopoulos2015}). The magnetic flux accumulated and amplified near SMBH is so strong that within a critical radius $R_{\rm m}$ \citep{Narayan2003}, the magnetic force is comparable to that of the gravity. The magnetic interchange instability makes the accretion operate, and the accreting gas moves inward in dense spirals (e.g., \citealt{Narayan2003, Tchekhovskoy2011, McKinney2012, White2019}). 

In this work, we adopt an analytic MAD model from \cite{Xie2019}, which assumes a non-spinning BH and adopts a pseudo-Newtonian gravitational potential. Following detailed general-relativistic magneto-hydrodynamic (MHD) simulations of MAD (e.g., \citealt{Tchekhovskoy2011, McKinney2012, White2019, Chatterjee2022}), this simplified model assumes that in the MAD regime, there are several (likely 2$-$4) dense gas spirals, surrounded by dilute but highly magnetized ``voids''. These gas spirals roughly reach a pressure equilibrium with surrounding magnetized voids. The details of our MAD model can be found in \citet{Xie2019}, with some updates in \citet{Xie2023}. Most importantly, unlike most simulations that adopt a single-fluid approximation, we use separate energy equations for the decoupled electrons and ions \citep{Narayan1995, Xie2019}. Below we highlight the main points.

The global vertical component of the magnetic field $B_z$ is fixed to follow a $R^{-1.1}$ profile, following simulations of \citet{McKinney2012}, and its strength is determined by the azimuthally-averaged plasma $\beta$ (gas-to-magnetic pressure ratio) parameter at the outer boundary $R_{\rm out} = 200\, R_{\rm g}$, i.e., $\bar{\beta}_{z0}=1$. Other global magnetic field components are determined through their ratio to $B_z$, i.e., $B_r/B_z$ and $B_\phi/B_z$, and they are constrained by, respectively, the magnetic Prandtl number $\mathcal{P}_{\rm m}=2$ and a parameter $\kappa_\phi = -0.5$. Such setup ensures that the vertical magnetic flux threading the accretion flow inside $10\,R_{\rm g}$ is about $50\,(\mdotbh R_{\rm g}^2 c)^{1/2}$, in agreement with that threading the BH reported in numerical simulations of MAD (\citealt{Tchekhovskoy2011, McKinney2012, Chatterjee2022}). The stress of global ordered magnetic fields, approximately proportional to $B_z B_\phi$  \citep{Xie2019}, can then be derived. We also include a turbulent field, by setting the plasma $\beta_{\rm t}=10$. As usual, we mimic the turbulent stress term, which is automatically (and implicitly) present in MHD simulations, via a radius-independent viscosity parameter $\alpha_{\rm vis} = 0.3$.

Numerical simulations suggest that hot accretion flows like MADs also have outflows (\citealt{Yang2021}, and references therein). In this work we adopt a smooth function of mass accretion rate as \citep{Xie2023},
\begin{equation}
\dot{M}(R) = \mdotbh \left[1+ \frac{(R-R_{\rm BH})^2}{R^2_{\rm flat}}\right]^{s/2},
\end{equation}
where $R_{\rm flat}=6 R_{\rm g}$ and $s=0.2$ are adopted from the non-spinning MAD case of \citet{Yang2021}. This expression ensures the suppression of outflow when $R\lesssim R_{\rm flat}$ and a power-law profile of $\dot{M}\propto R^s$ when $R\gg R_{\rm flat}$.

Because the optical depth of MAD is low, the radiative mechanisms are synchrotron, bremsstrahlung, and (inverse) Compton scattering \citep{Narayan1995, Manmoto1997, Xie2019}. We typically (e.g., in Sec.\ \ref{sec:m104case}) assume that the hot electrons in MAD follow a relativistic thermal distribution, $n_{\rm th}(\gamma)\propto \frac{\gamma^2\beta}{\theta_{\rm e} K_2(1/\theta_{\rm e})}\,\exp(-\gamma/\theta_{\rm e})$, where $\theta_{\rm e}=kT_{\rm e}/m_{\rm e} c^2$ and $K_2(x)$ is the modified Bessel function of the second kind. Considering particle acceleration by magnetic reconnection, we also extend to consider a small fraction of electrons in MAD to follow a power-law (PL) distribution, i.e., $n_{\rm pl}(\gamma) \propto \gamma^{-p_{\rm mad}}$, where $\gamma m_e c^2$ is the electron energy. The intrinsic/initial distribution index is fixed to $p_{\rm mad} = 2.5$. The total thermal energy of these PL electrons is only a small fraction $\xi_{\rm pl, mad}$ (note that $\xi_{\rm pl, mad}\ll 1$) compared to the thermal energy of thermal electrons. The minimum energy of PL electrons is fixed to $\gamma_{\rm pl,min}^{\rm mad}=2$. The maximum energy can be evaluated by  balancing the particle acceleration rate to synchrotron cooling rate \citep{Zdziarski2014}, i.e. $\gamma_{\rm pl, max}^{\rm mad} = \sqrt{\frac{9 \eta_{\rm acc} B_{\rm cr}}{4 \alpha_{\rm f} B}}$. Here $\alpha_{\rm f} = 2 \pi e^2 / (h c)\approx 1/137.0$ is the fine-structure constant and $B_{\rm cr}=2\pi m_e^2 c^3/(eh)\approx 4.41\times10^{13}$ Gauss is the critical magnetic field constant (also named Schwinger limit). We set the acceleration efficiency $\eta_{\rm acc}=1$. Note that non-thermal electrons may still exist above $\gamma_{\rm pl, max}$, but their population is further reduced exponentially \citep{Zdziarski2014, Kimura2020, Kuze2024}. In this work, we for simplicity ignore their impact, which is an good approximation for electrons as $p_{\rm mad} > 2$ and $\gamma_{\rm pl, max}^{\rm mad}\sim 10^{7-8}$.

More energetic electrons (larger in $\gamma$) suffer much stronger radiative cooling. Consequently, we can define a ``cooling-break Lorentz factor'' $\gamma_{\rm cooling}$, above which PL electrons follow a steeper profile of $n_{\rm pl}(\gamma)\propto \gamma^{-(p_{{\rm mad}}+1)}$ when a steady state is achieved \citep{Rybicki1979}. Technically $\gamma_{\rm cooling}$ is determined by equaling the radiative cooling timescale to the accretion timescale, 
\begin{equation}
    t_{\rm cooling} \equiv \frac{3}{4} \frac{8 \pi m_{\rm e} c}{\sigma_{\mathrm{T}} \gamma_{\rm cooling} \beta_{\rm e}^{2} B^{2}} = t_{\rm {acc }} \equiv \frac{R}{|V_{\rm R}|},
    \label{eq:tcool}
\end{equation}
and it reads (electron velocity is approximated as $\beta_{\rm e} \approx 1$),
\begin{equation}
    \gamma_{\rm cooling} = \frac{6\pi m_{\rm e} c}{\sigma_{\rm T}} \cdot \frac{|V_{\rm R}|}{B^2R}
    \label{eq:gamma_cooling}
\end{equation}
All the quantities are of their usual meanings. 

\begin{figure*}
  \includegraphics[width=0.48\textwidth]{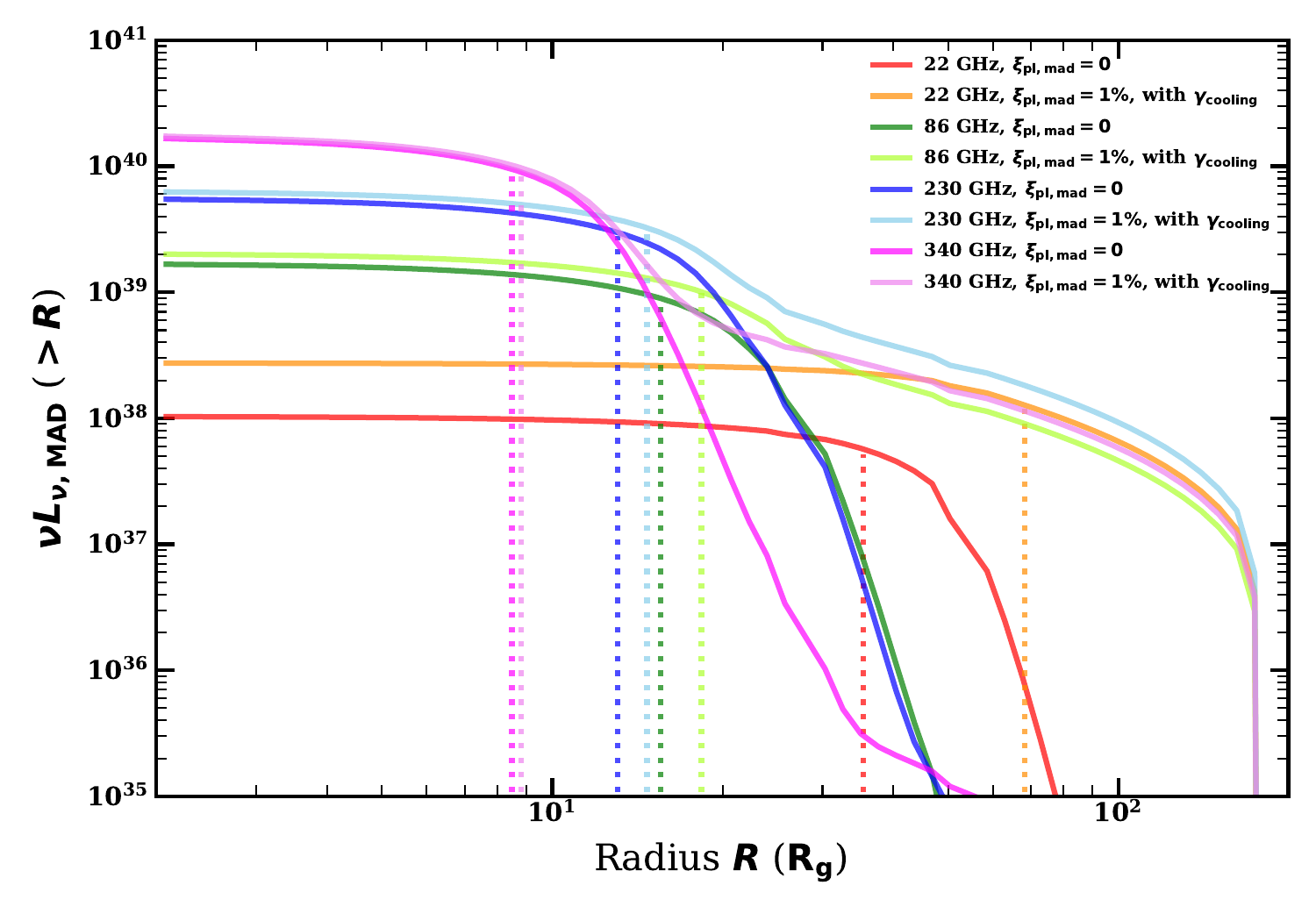}
  \hspace{0.3cm}\includegraphics[width=0.48\textwidth]{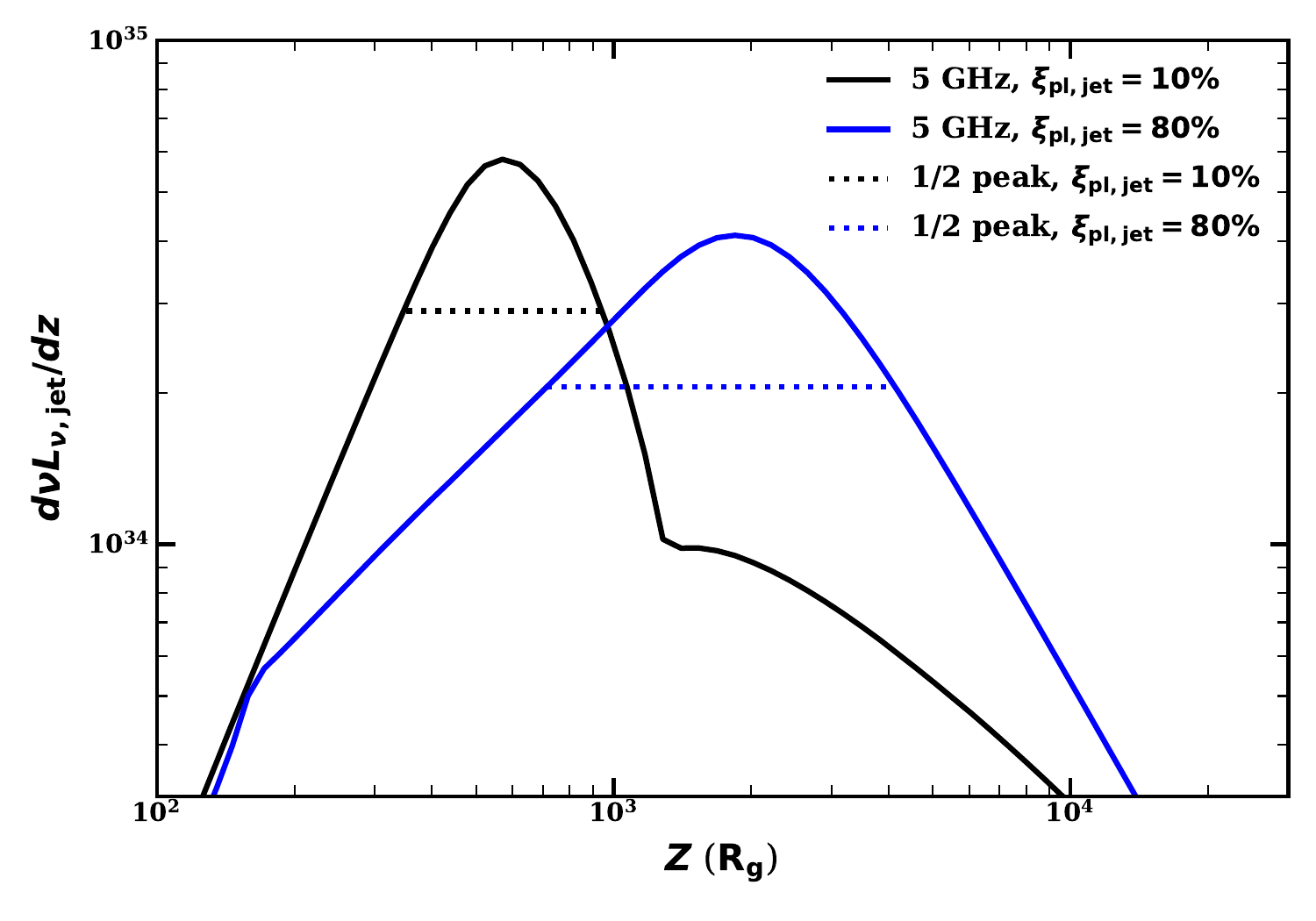} 
  \caption{\textit{Left panel:} Cumulative monochromatic radiation distribution $\nu L_{\nu, MAD}(>R)$ (from infinity to radius $R$) of MAD, as a function of radius $R$. The BH mass is $10^9\,\msun$ and the accretion rate at $200\,R_{\rm g}$ is $\dot{m}_0 = 1\times10^{-4}$. As specified in color, several radio frequencies are considered, i.e., 22, 86, 230, 340 GHz. Two different electron populations are investigated, one for purely thermal electrons ($\xi_{\rm pl, mad}=0$), and the other for $1\%$ non-thermal electrons (radiative cooling to those PL electrons considered self-consistently) ($\xi_{\rm pl, mad}=1\%$). The dotted line with the same color marks the model-predicted radius of the MAD as observed at that frequency. Clearly, the radio size of MAD decreases with frequency, and a small fraction of PL electrons only slightly increases the radio size. \textit{Right panel:} Radiation distribution at 5 GHz along the jet direction $z$. Here the mass loss rate and the bulk Lorentz factor of the jet are respectively, $\dot{m}_{\rm jet} = 8.1\times 10^{-7}$ and $\Gamma_{\rm jet} = 10$. Two different electron populations are considered, i.e., $\xi_{\rm pl, jet}=10\%$ (black solid) and $80\%$ (blue solid). Theoretically the jet location in radio is defined at where it reaches its peak value, while the jet size is evaluated by the horizontal dotted line that intersects 1/2 of the peak value.}
  \label{fig:size}
\end{figure*}

\subsection{Jet} \label{sec:jet}

The formation (launch, acceleration/deceleration, collimation, etc) and composition (e.g., $e^-$-$e^+$ leptonic versus $e^-$-$p$ hadronic) of relativistic jets in black hole accretion systems is a long-standing unsolved problem. Theoretically speaking, a jet can be formed by extracting the rotation energy of the BH (\citealt{Blandford1977}; so-called the ``BZ-jet'') or the accretion flow \citep{Blandford1982}. The dynamics of a relativistic jet has been investigated extensively through MHD simulations (e.g., \citealt{Tchekhovskoy2011}), and these jets are primarily composed of electromagnetic Poynting-flux. Most of these studies focus on a BZ-jet. Recently, important progress has been made through a detailed comparison between observations and simulations of the inner $\lesssim 500 R_{\rm g}$ regions of the jet in M87 \citep{Yang2024}, where they confirm the BZ-jet picture and further identify magnetic reconnection as the key mechanism to accelerate electrons.

In this work, an internal shock prescription jet model \citep{Spada2001,Yuan2005} is adopted, which has been applied to $\gamma$-ray bursts and radio-loud AGNs. The dynamical structure of the jet is similar to the original BZ-jet model, but with several modifications \citep{Yuan2005, Xie2016}. First, we limit ourselves to regions after ``mass loading'', i.e., we assume the jet is composed of normal plasma (protons and electrons). We note that our jet model can be extended to a leptonic one easily. Secondly, we omit for simplicity the acceleration or deceleration of the jet (e.g., as observed in the jets of several nearby LLAGNs such as M87, \citealt{Asada2012}). Finally, we ignore jet collimation and adopt a conical jet morphology with the half-opening angle fixed to $\theta_{\rm jet}=0.1$ (equivalently $\theta_{\rm jet}=5.73\degr$).

There are several additional model parameters of the jet, which we follow \citet{Xie2016}. The primary parameter is the mass loss rate into the jet $\dot{M}_{\rm jet}$, or in Eddington units, $\dot{m}_{\rm jet} \equiv \dot{M}_{\rm jet}/\dot{M}_{\rm Edd}$.\footnote{The Eddington mass accretion rate is defined as $\dot{M}_{\rm Edd} = 10 L_{\rm Edd}/c^2$, where the Eddington luminosity for accretion onto a $\mbh$ BH is $L_{\rm Edd} = 1.3\times10^{47}\,{\rm erg\,s^{-1}}\, (\mbh/10^9\msun)$.} We assume a fraction of $\xi_{\rm pl,jet}$ electrons are accelerated into a PL distribution, where the injected distribution index is set to $p_{\rm jet} =2.2$ (Again, $n_{\rm jet,pl}(\gamma)\propto \gamma^{-p_{\rm jet}}$; see Sec. \ref{sec:mad} for a similar value). Unlike the mainbody of MAD, here we assume that most of the electrons in jet are non-thermal and take $\xi_{\rm pl,jet}=80\%$ as the fiducial value. The radiative cooling to PL electrons at high-$\gamma$ end is also included, where we replace the accretion timescale of Equation (\ref{eq:tcool}) with the dynamical timescale of the jet $t_{\rm dyn} = z/V_{\rm jet}$. Additionally, the energy densities of the accelerated electrons and the amplified magnetic field are determined by two parameters, $\epsilon_{\rm e} = 0.02$ and $\epsilon_B = 0.02$, which describe the fractions of the shock energy that was transferred to electrons and the magnetic field, respectively. The impact of different values of $\epsilon_{\rm e}$ and $\epsilon_B$ on the outcome jet spectrum is discussed in \citet{Xie2014}. Finally, we note that the maximal energy of PL electrons is evaluated as the MAD case (see Sec. \ref{sec:mad}), and the minimal energy is determined based on the definition of $\xi_{\rm pl,jet}$ \citep{Spada2001, Yuan2005}, where $\gamma_{\rm pl,min}^{\rm jet} \propto (\epsilon_{\rm e}/\xi_{\rm pl,jet})$.

We caution that, for LLAGNs, at distances $\ga 10$ kpc away from the central SMBH, there usually exist two large-scale giant radio lobes. They are generated by interactions between the relativistic jet and the ambient gas in and near the host galaxy. Observationally, these giant radio lobes can easily be spatially resolved by current facilities. In this work we do not include their emission.

\subsection{Coupling between MAD and jet}\label{sec:coupling}

The coupling between accretion (i.e., MAD) and ejection (e.g., jet or outflow) still cannot be determined from first principles; we  thus rely on observational constraints. It has been discovered that, in AGNs and black hole binaries, the monochromatic radio luminosity at 5 GHz or 8.5 GHz $L_{\rm R}$ (defined as $L_{\rm R}\equiv\nu L_{\rm \nu}$), the integrated  2--10 keV X-ray luminosity $L_{\rm X}$ (defined as $L_{\rm X} \equiv \int_{\rm 2\,keV}^{\rm 10\,keV} L_{\rm \epsilon} d\epsilon$, where $\epsilon = h\nu$ is the photon energy), and the BH mass $\mbh$, have a tight relationship. In the logarithmic space of ($L_{\rm R}$, $L_{\rm X}$, $\mbh$), they form a so-called Fundamental Plane (FP; \citealt{Merloni2003, Falcke2004}. Hereafter M03 FP) of BH activity, i.e., $\log L_{\rm R} = 0.6\log L_{\rm X} + 0.78\log \mbh + const$. The FP has been investigated statistically based on samples selected under various methods (e.g., \citealt{Merloni2003, Falcke2004, Gultekin2009, Xie2017}), and by case studies as well (For individual LLAGNs, see e.g., \citealt{Bell2011, King2011, King2013, Xie2016, Jones2017}). 

Theoretically, low-frequency radio emission mainly comes from relativistic jets, while the X-ray emission mainly originates from hot accretion flow (e.g., \citealt{Ho2008, Yuan2014, Xie2016a}). The FP provides strong evidence for the coupling between hot accretion flow (MAD here) and jet. 

In this work, for a given BH mass $\mbh$ and an accretion power (through $\dot{m}_0$, or equivalently $L_{\rm X}$), we manually adjust the jet power (through $\dot{m}_{\rm jet}$, since $\Gamma_{\rm jet}$ is fixed), to set the MAD-jet coupling. During this process, we take the M03 FP as a reference.

\begin{figure*}
  \centering
  \includegraphics[width=0.48\textwidth]{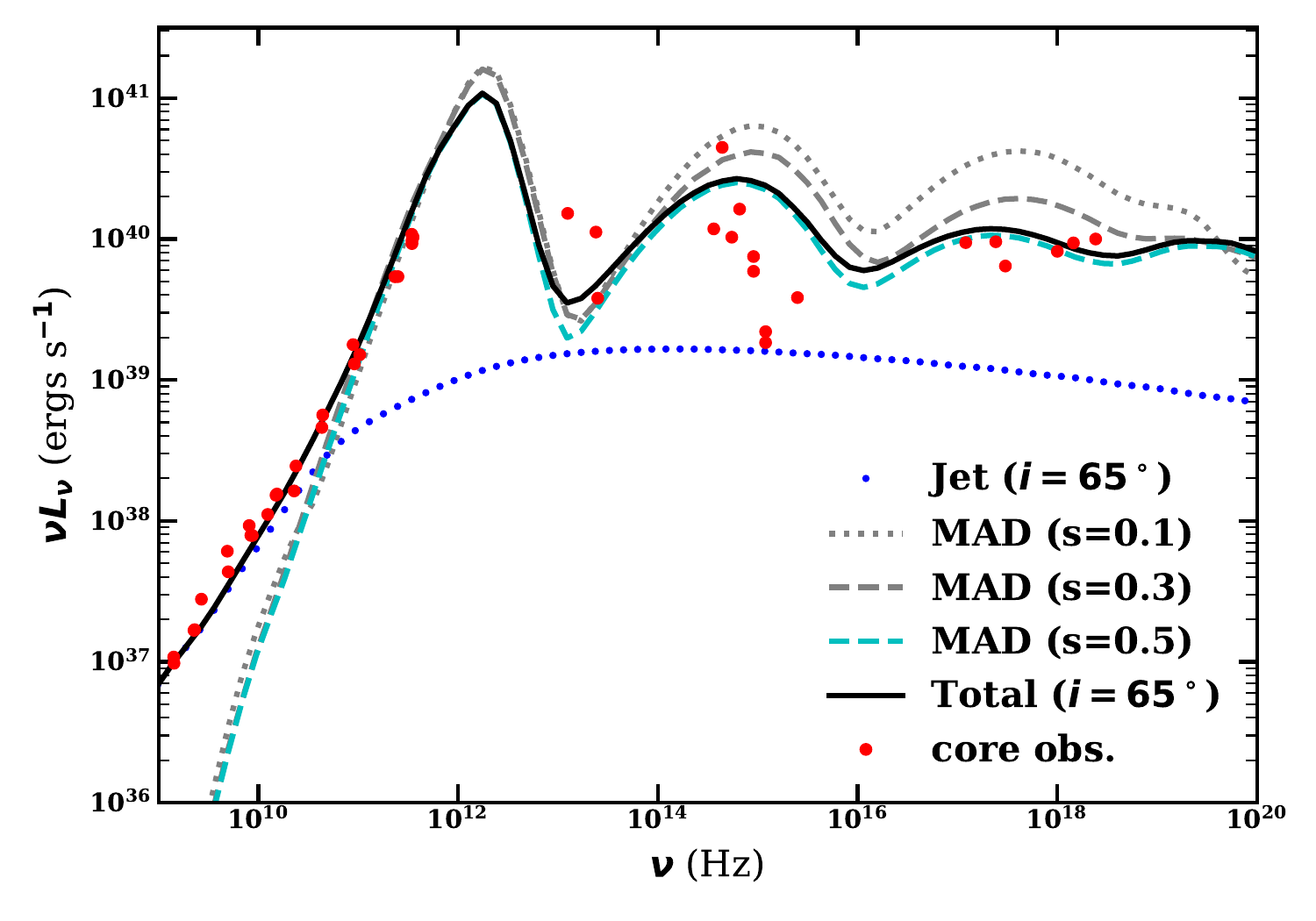}
  \includegraphics[width=0.48\textwidth]{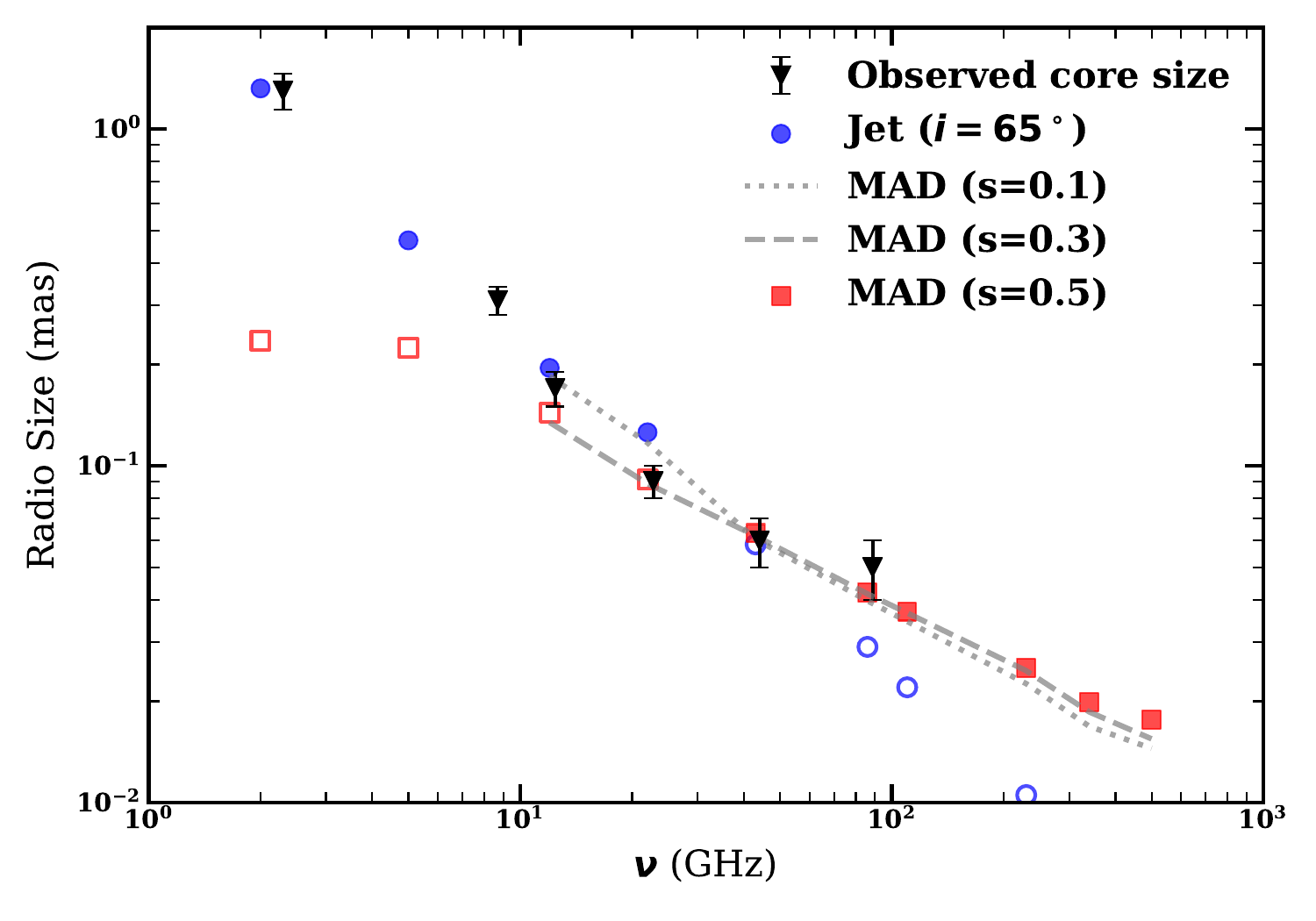}
  \caption{Theoretical modelling of M104 under the MAD.th-jet model. For simplicity, we only show cases with viewing-angle $i=65\degr$. {\it Left panel}: broadband SED. Here the red dots are gathered by \citet{Yan2024}. The cyan dashed, the blue dotted, and the black solid lines represent emission from MAD (outflow parameter $s=0.5$), jet, and MAD+jet (total), respectively. {\it Right panel}: the radio core size (note that $R_{\rm g}\approx 1\times10^{-3}$ mas). Here the black triangles (with error-bars) are from high-resolution observations \citep{Yan2024}. Theoretical results from MAD and jet are shown by red squares and blue circles, respectively. For clarity, they are shown in filled or open markers, depending on their flux-dominance. As can be seen from left panel, the jet (MAD) radiation dominates below (above) $\sim 30$-40 GHz. In both panels, we also show MAD.th calculations with $s=0.1$ (gray dotted curve) and $s=0.3$ (gray dashed curve), for comparison. Accretion rates are adjust to fit the radio luminosity.}
  \label{fig:m104fit} 
\end{figure*}

\subsection{Observed Location and Size of LLAGNs}

Even if we focus on the spatially-resolved nuclear (sub-kpc or even more compact) regions of LLAGNs, their emission still has several origins, e.g., most of the infrared emission may come from the dusty torus (e.g., \citealt{Urry1995}). Here we further limit ourselves to the central BH accretion system. In this case, the emission may originate from either the accretion flow (see Sec.\ \ref{sec:mad}) or the jet (see Sec.\ \ref{sec:jet}), which is distinctive from each other in radiative properties. Consequently, the location and the size of LLAGNs vary with luminosity and observed wavebands.

The location and the size are evaluated separately for the MAD and the jet. Observationally the MAD appears as a quasi-spherical ring, with its center remaining dark (flux highly suppressed) due general-relativistic effect of SMBH there (see \citealt{EHT2019, EHT2022}). Neglecting the relativistic Doppler effects, the central site of the MAD emission is approximately at SMBH ($a_{\rm MAD} = 0$), and is independent with frequency $\nu$ (photon energy $h\nu$). The magnetic field, density and the gas velocity varies at different radius in MAD. Consequently, the size of radio emission also varies with observed frequency/waveband (for the synchrotron emission in the weak-field version of hot accretion flow, see \citealt{Narayan1998}). As will be shown later in the left panel of Figure \ref{fig:size}, we follow the spirit of baseline-associated VLBI technique and calculate the emission outside $R$ as (e.g., \citealt{BandX2019}),
\begin{equation}
\nu L_{\nu,\,{\rm MAD}}(>R) = \int_{R}^{\infty} \int\nu I_\nu(R,\phi) d\phi RdR.
\end{equation}
Here $I_\nu(R,\phi)$ defines the local radiative intensity per unit area as observed by a distant observer. The size of MAD $d_{\rm MAD}$ can be estimated by the diameter of the MAD ring, whose outer boundary is defined by the half maximum of the total radiation (shown in this plot as vertical dotted lines with the same color). We note that emission from inside the photon ring radius of SMBH is highly suppressed and becomes negligible to distant observer, which explains the flat shape of  $\nu L_{\nu,\,{\rm MAD}}$ at $R<4-5R_{\rm g}$. 

The radio size of jet is primarily dominated by its spread along the jet direction, but not perpendicular to it \citep{Xie2016}. We thus evaluate its size $d_{\rm jet}$ as the full-width-at-half-maximum (FWHM) of the $d(\nu L_{\nu,\,{\rm jet}})/dz - z$ curve, which is shown by the right panel of Figure \ref{fig:size}. The radio jet (at a given frequency) locates  where it reaches its peak value, and is denoted as $a_{\rm jet}$. 

For a given observed frequency, when the emission from MAD far exceeds that from jet, i.e. $L_{\nu,\,{\rm MAD}} \gg L_{\nu,\,{\rm jet}}$, the location and the size will be determined by MAD, and vice versa. The situation becomes complicated when $L_{\nu,\,{\rm MAD}} \sim L_{\nu,\,{\rm jet}}$. In this case, if the observation has sufficient spatial resolution and flux sensitivity, we should observe two isolated components, and the separation of these two components depends on the observed frequency (Note that $a_{\rm jet} - a_{\rm MAD} = a_{\rm jet}$. See Figure~\ref{fig:m104fit} and Figure~\ref{fig:size_mad.th_jet} below).

\section{A Case study: nearby LLAGN M104} \label{sec:m104case}

\begin{table}
    \centering
    \caption{Parameters of the MAD-jet Model in M104}
    \label{tab:m104fit}
    \resizebox{\columnwidth}{!}{
    \begin{tabular}{lll}
        \hline
        \textbf{Parameter} &\textbf{Value} & \textbf{Definition/Note} \\
        \hline
        \multicolumn{3}{c}{\textbf{M104 System}} \\
        \hline
        $\mbh$ & $1\times10^9\msun$ & black hole mass\\
        $\theta_{\rm obs}$ & $65\degr$ & viewing angle/inclination\\
        \hline
        \multicolumn{3}{c}{\textbf{MAD}} \\
        \hline
        $\dot{m}_0$ & $(6.5/8.5/11) \times 10^{-4}$ & $ \dot{M}/\dot{M}_{\rm Edd}$ at boundary $200\,R_{\rm g} $ \\
        $s$ & $0.1/0.3/0.5$ & Outflow parameter \\
        $\alpha_{\rm vis}$ & $0.3$ & Viscosity parameter \\
        $\beta_{\rm turb}$ & $10.0$ & $P_{\rm gas}/P_{\rm mag}$ \\
        $\delta$ & $0.1$ & Fraction of electron viscous heating\\
        $\xi_{\rm pl, mad}$ & $0\%$ & Energy fraction of power-law electrons\\
        $p_{\rm pl, mad}$ & $2.2$ & Energy distribution index of power-law electrons\\
        $\beta_{z_0}$ & $1.0$ & $\beta_{z_0} = C_{\rm arm} \cdot \frac{8\pi\,P_{\rm gas}}{B_{z,\rm ave}^2}$ at boundary $ 200\,R_{\rm g}$ \\
        $s_{b_z}$ & $1.1$ & $B_z \sim r^{-s_{b_z}}$ \\
        $Pr_{m}$ & $2.0$ & Used in $\frac{B_r}{B_z} = \frac{1}{Pr_m}\,\frac{H}{R}$ \\
        $\kappa_{\phi, 0}$ & $-0.5$ & $\frac{B_\phi}{B_z} = \kappa_{\phi 0}\,\frac{V_\phi}{|V_r| + V_K}\,\frac{H}{R}$\\
        \hline
        \multicolumn{3}{c}{\textbf{Jet}} \\
        \hline
        $\dot{m}_{\rm jet}$ & $1.5 \times 10^{-6}$ & Mass loss rate of jet \\
        $\phi_{\rm jet}$ & $5.73\degr$ & Intrinsic half-opening angle, equivalently $0.1$ radian \\
        $\Gamma_{\rm jet}$ & $1.02$ & bulk Lorentz factor. Typical value of LLAGNs: $\Gamma_{\rm jet}=10$ \\
        $\xi_{\rm pl, jet}$ & $80\%$ & Energy fraction of power-law electrons.\\
        $p_{\rm pl, jet}$ & $2.2$ & Energy distribution index of power-law electrons\\
        $\epsilon_{\rm e}$ & $0.02$ & Acceleration-to-shock energy density ratio \\
        $\epsilon_B$ & $0.02$ & Magnetic-to-shock energy density ratio \\
        \hline
    \end{tabular}
    }
\end{table}

We take the nucleus of the nearby (located at a distance $d\approx$10 Mpc) early-type spiral galaxy, M104 (also named Sombrero galaxy and NGC 4594), as an testbed for the MAD-jet model. The host galaxy has an inclination of $84\degr$ \citep{Menezes2013, Menezes2015, Sutter2022} and the mass of the central SMBH is $\sim1\times10^9\msun$ \citep{Menezes2015}. 

Similar to several nearby LLAGNs like Sgr A* \citep{Bower2015}, M87 \citep{Algaba2024} and M84 \citep{BandX2019}, the nucleus of M104 also has an extensive set of high-resolution radio observations spanning from $\sim$2 GHz up to $\sim$90 GHz \citep{Eracleous2010, Fern'andez-Ontiveros2023, Yan2024}. Such broad and intense coverage in frequency is crucial for our investigation. As noted by \citet{Yan2024}, the broadband radio spectrum of M104 reveals a shift in spectral index at frequencies below and above $\sim 40$ GHz. The frequency-dependence of the VLBI core size across frequencies ranging from 2.3 to 88 GHz was also reported by \citet{Yan2024}, in which below $\sim 30-40$ GHz a power-law relation $d_{\rm core}\sim \nu^{-1.13\pm0.04}$ was reported \citep{Hada2013, Yan2024}. A deviation to the extrapolation of this power-law relation was noted at 86 GHz \citep{Yan2024}.

The above two observational properties of M104 have been understood under a coupled hot accretion flow-jet framework  (\citealt{Yan2024}, see also \citealt{BandX2019}), where the hot accretion flow was a conventional weak-field one \citep{Narayan1994, Narayan1995}. We also note that jet emission is actually not modeled in \citet{Yan2024}. 

In this section, we update their work to fully model LLAGN M104 under the MAD.th-jet scenario (`.th' represents purely thermal electrons, see Sec. \ref{sec:size_LLAGN_MAD.th} below). The model parameters are listed in Table \ref{tab:m104fit}, and the results are shown in Figure \ref{fig:m104fit}, where the left panel shows the broad-band spectrum and the right panel shows
the radio core size. The observational data (shown by red filled circles in the left panel, and black triangles in the right panel) are gathered from literature \citep{Eracleous2010, Fern'andez-Ontiveros2023, Yan2024}. We note that the inclination of the accretion flow in M104 remains inconclusive, either $\theta_{\rm obs}\lesssim25\degr$ \citep{Hada2013} or $\theta_{\rm obs}\sim66\degr$ \citep{Menezes2015, Yan2024}. In this work, we take $\theta_{\rm obs}=65\degr$. We notice that, the accretion rate (bolometric luminosity) of M104 is moderately high. In this case, even if we assume there may intrinsically create a notable fraction of non-thermal electrons through magnetic reconnection in the MAD region, these non-thermal electrons will cool down efficiently (see Equation (\ref{eq:tcool}) above and Sec.\ \ref{sec:size_LLAGN_MAD.pl}). Thus our MAD model only considers thermal electrons (i.e. MAD.th).

One notable exception of M104 is that, unlike typical radio-loud AGNs whose $\Gamma_{\rm jet} \approx 10$ (see e.g., \citealt{Lister2009}), the bulk velocity of the jet in M104 is considerably low \citep{Hada2013, Yan2024}, i.e., $V_{\rm jet}\approx 0.2\,c$. We thus fix the Lorentz factor as $\Gamma_{\rm jet} = (1-V^2_{\rm jet}/c^2)^{-1/2} = 1.02$. Consequently, the relativistic beaming effect is highly suppressed in M104. We find that it is difficult to constrain $\xi_{\rm pl, jet}$. With our own preference, we adopt $\xi_{\rm pl, jet}=80\%$ (see Figure \ref{fig:size_mad.th_jet} below.)

As shown in Fig. \ref{fig:m104fit}, our MAD.th-jet model agrees with the radio data up to $\sim$ 90 GHz, and the X-ray data in general. Besides, the radio size is also nicely explained (see also \citealt{Yan2024}). We find that, compared to self-absorbed synchrotron (high-frequency radio), the Compton scattering (evident in X-rays) is more sensitive to $\dot{M}$ (or equivalently, optical depth; \citealt{Narayan1995}). For completeness, we also show in this figure two additional choices of $s$, where the $\dot{M}$ is adjusted to fit the radio data. Clearly they differ by a factor of $2$-3 in X-rays. We suggest that in future a detailed observations of SED in high-frequency radio (upto THz and far infrared) and X-rays can help to break the degeneracy in accretion rate $\dot{M}$ and outflow parameter $s$ in AGNs.

\section{Radio Core Size of LLAGNs}
\label{sec:radio_size}

Now, we extend our investigations to LLAGNs in local Universe at different luminosities (and different accretion rates). For this investigation, the BH mass is fixed to $\mbh = 10^9\msun$, and the inclination of the system is set to $\theta_{\rm obs}=45\degr$. We take a fixed bulk Lorentz factor of the jet as $\Gamma_{\rm jet}=10$, which is a typical value of radio-loud AGNs (e.g., \citealt{Lister2009}). We emphasize that outflow in LLAGNs is poorly constrained, mostly because they are totally ionized (with temperature $\sim10^{7-9}$K) thus do not show any emission/absorption lines. For our general investigation, the outflow parameter is set to $s=0.1$, which is more commonly adopted in LLAGNs compared to $s=0.5$ as suggested by M104. We focus on changes in mass accretion rate $\dot{m}_0$. The coupling between accretion (MAD here) and ejection (jet) is assumed to follow the M03 FP. Obviously LLAGNs that deviate from the M03 FP can also be investigated accordingly. Values of other  microphysics parameters (except $\xi_{\rm pl,\,mad}$) take those of M104, see Table \ref{tab:m104fit}.

\begin{figure}
\centering
\includegraphics[width=0.95\columnwidth]{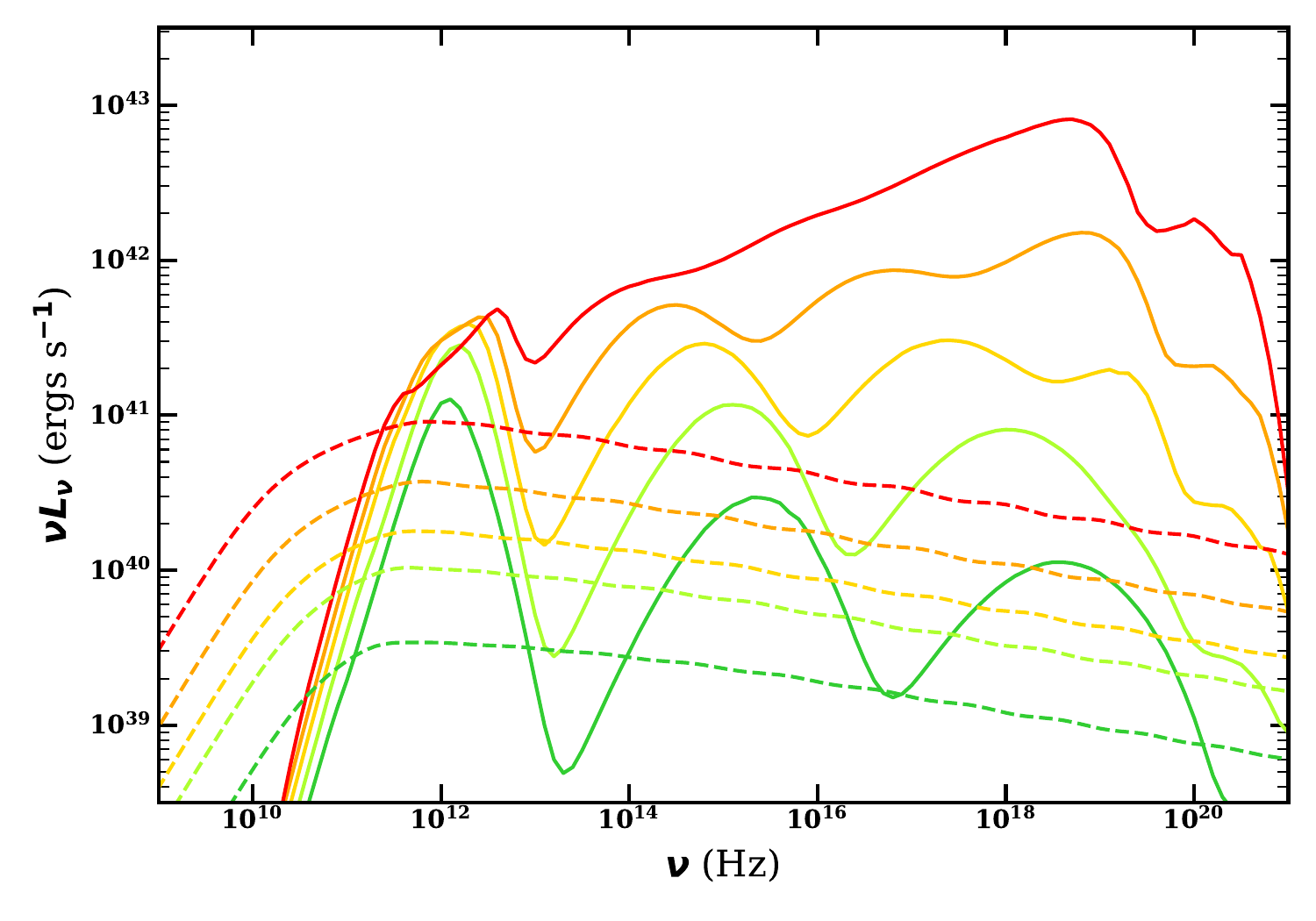}
  \caption{Spectrum of LLAGNs under the MAD.th-jet model (only thermal electrons in MAD). The emission from MAD.th is shown in solid curves, where from top to bottom, are results for $\dot{m}_0=1\times10^{-2}$, $3.16\times10^{-3}$, $1\times10^{-3}$, $3.16\times10^{-4}$, $1\times10^{-4}$, respectively. The dashed lines with the same color represent the jet emission that follows the M03 FP.}
  \label{fig:SEDth}
\end{figure}

\subsection{Fiducial MAD.th Case: MAD that Only Consists Thermal Electrons} \label{sec:size_LLAGN_MAD.th}

We first consider the case that there are only (relativistic) thermal electrons within the MAD (MAD.th hereafter). For the reasons addressed later in Sec.\ \ref{sec:size_LLAGN_MAD.pl}, we take this as the fiducial case.

We first show in Figure\ \ref{fig:SEDth} broadband SED ranging from $10^9$ Hz up to $10^{21}$ Hz. The solid curves from top to bottom, are MAD radiation under mass accretion rates (at boundary $200\ R_{\rm g}$) of $\dot{m}_0=1\times10^{-2}$, $3.16\times10^{-3}$, $1\times10^{-3}$, $3.16\times10^{-4}$, and $\dot{m}_0 =1\times10^{-4}$. It is well-known that, at low accretion rates, the spectrum of MAD (and hot accretion flows in general) exhibits three distinct peaks/bumps (e.g., \citealt{Narayan1995,Xie2019}), the $10^{11-13}$ Hz bump is the self-absorbed synchrotron radiation, the $10^{13-15}$ Hz bump is the first-order Compton scattering of the synchrotron radiation by the hot electrons in MAD, and the $10^{16-19}$ Hz bump represents a combination of high-order Comptonization and bremsstrahlung.\footnote{Bremsstrahlung is insignificant in our calculations. The bremsstrahlung emission, roughly proportional to $n^2 R^3 T_{\rm e}^{-1/2}\exp(-h\nu/kT_{\rm e})$, is most observable at outer $R>10^3\ R_{\rm g}$ regions of the hot accretion system.} As accretion rate (optical depth) increases, Compton scattering becomes more and more evident, and the X-ray emission will become a power-law. This is clearly shown in this plot when $\dot{m}_0 \ga 3\times10^{-3}$.

\begin{figure}
\centering
\includegraphics[width=0.95\columnwidth]{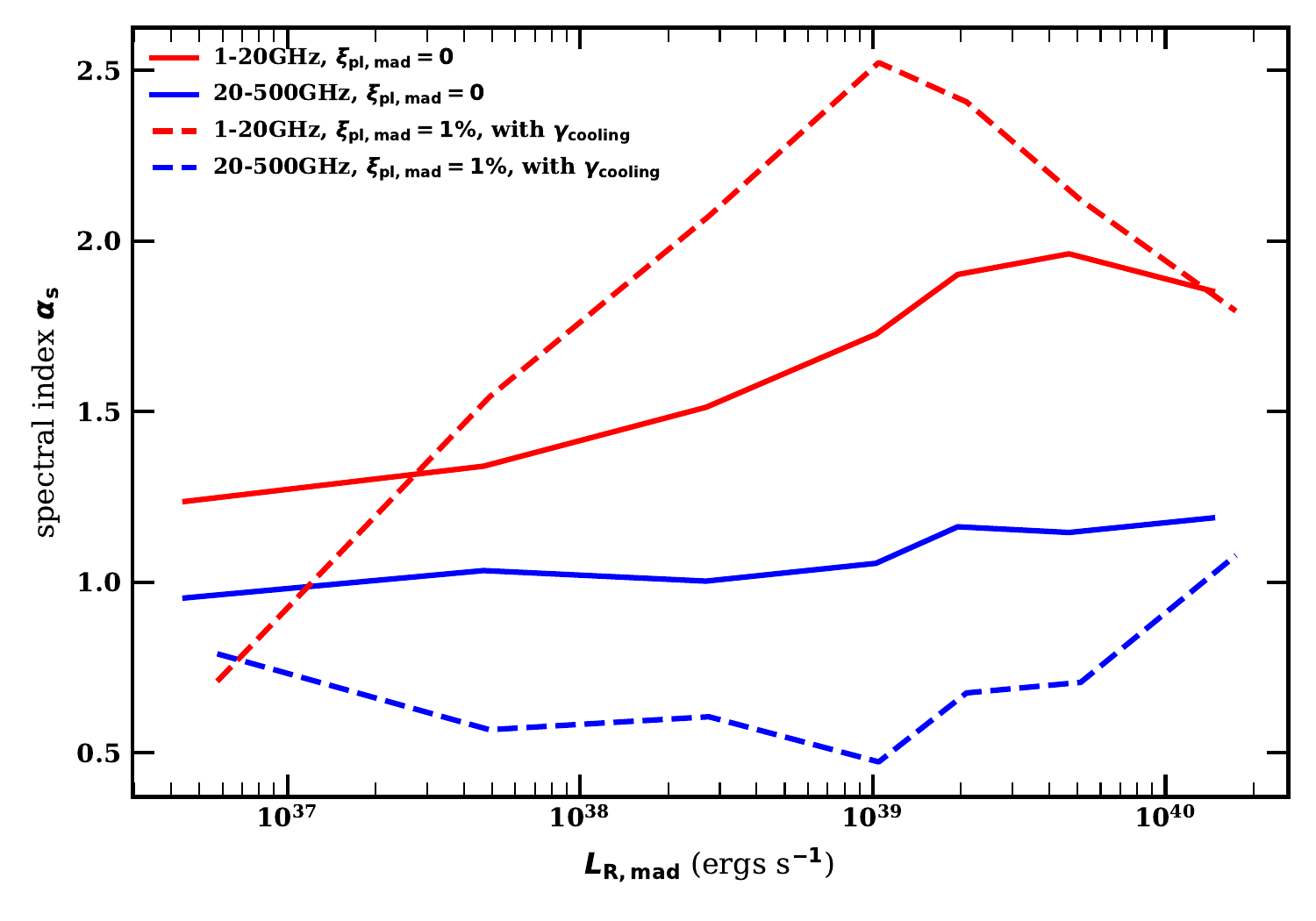}
  \caption{Spectral index of the radio emission of MAD $\alpha_{\rm s}$ as a function of 5 GHz radio luminosity of MAD $L_{\rm R,\,mad}$(caution that jet emission at this frequency is not included). The red and blue curves, represent $\alpha_{\rm s}$ derived at 1-20 GHz and 20-500 GHz, respectively. The solid and the dashed property of each curve define respectively, MAD.th case with $\xi_{\rm pl,\,mad}=0$ and self-consistent MAD.pl case with $\xi_{\rm pl,\,mad}=1\%$ and $\gamma_{\rm cooling}$.}
  \label{fig:mad_alpha_s}
\end{figure}

For a closer look at the synchrotron radio emission of the MAD.th model, we then study its spectral index $\alpha_{\rm s}$ (defined as $L_\nu \propto \nu^{\alpha_{\rm s}}$). For this purpose, we separate the 1-500 GHz into two bands, i.e., the low-frequency band of 1-20 GHz, and the high-frequency band of 20-500 GHz. Above 500 GHz it will gradually enter into the optically-thin regime of synchrotron emission and reaches the peak synchrotron flux; below 1 GHz the MAD emission is admittedly faint and far below the jet emission. The results are shown by solid curves in Figure \ref{fig:mad_alpha_s}, where the X-axis represents the MAD luminosity at 5 GHz $L_{\rm R,\,mad}$ (cautioning that the jet emission at the same frequency is not taken into account). The red and blue colors indicate the $\alpha_{\rm s}$ in 1-20 GHz and 20-500 GHz, respectively. We find that, $\alpha_{\rm s}$ of the MAD spectrum of 20-500 GHz band is systematically lower than that of 1-20 GHz band. Besides, as accretion rate (and luminosity $L_{\rm R,\,mad}$) increases, we observe a gradual increase in $\alpha_{\rm s}$ (except the that of 1-20 GHz at high $L_{\rm R,\,mad}$), i.e., as the radio luminosity increases from $6\times10^{36}\,\ergs$ to $1.7\times10^{40}\,\ergs$, the spectral index $\alpha_{\rm s}(20-500\,{\rm GHz})$ increases from $\approx 1$ to $\approx 1.4$. All these reflect the dynamical changes of MAD due to radiative cooling \citep{Xie2019}. Finally we note that $\alpha_{\rm s}$ derived here is much larger (thus spectrum much steeper) than $\alpha_{\rm s}\approx 0.4$ reported in \citet{Mahadevan1997}, which is derived based on a self-similar solution of the conventional weak-field hot accretion flow at low accretion rates.

We now consider the jet emission. The dashed curves with the same color in Figure\ \ref{fig:SEDth} also shows the jet emission that follows the M03 FP (at the given accretion power $\dot{m}_0$). With a comparison of the emission between MAD.th and jet, we can find that, in the radio band (below $10^{12}$ Hz), the jet emission typically shows a curved self-absorbed pattern, while the MAD emission typically follows a steep spectrum. We note that such a shift in the spectral shape has been noted for decades, and the change in the radio origin has been proposed, i.e., the high-frequency radio emission comes from hot accretion flow, while low-frequency one originates from either jet in radio-loud systems, or PL electrons of hot accretion flows in some radio-quiet systems (e.g., \citealt{Yuan2003, LiuWu2013, BandX2019} and references therein). 

\begin{figure}
\centering
\includegraphics[width=0.95\columnwidth]{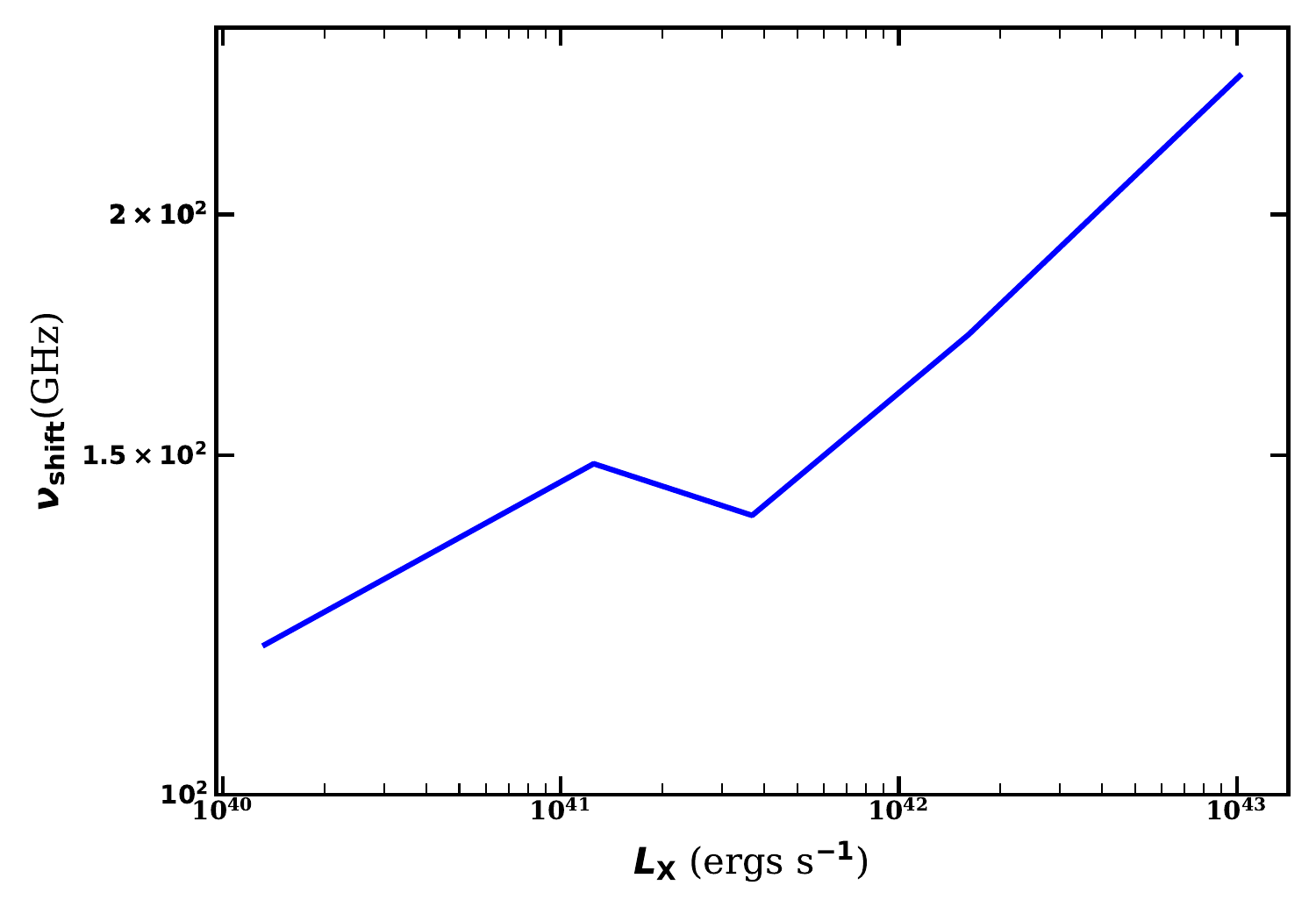}
  \caption{Evolution of the ``shift''/critical frequency $\nu_{\rm shift}$ (below which the jet emission dominates) as a function of the 2-10 keV X-ray luminosity. The BH mass of the LLAGN is assumed to be $10^9\msun$.}
  \label{fig:nu_shift}
\end{figure}

In order to characterize the change between these two origins, we here crudely define the ``shift'' frequency $\nu_{\rm shift}$, below (above) which the jet (MAD) emission exceeds the other. Based on the spectrum illustrated in Figure\ \ref{fig:SEDth}, we then show in Figure\ \ref{fig:nu_shift} the evolution of $\nu_{\rm shift}$ as a function of X-ray luminosity. As expected, we observe a gradual (and weak) increase in $\nu_{\rm shift}$ as X-ray luminosity (and $\dot{m}_0$) increases, i.e., for a factor of $\sim 10^3$ increase in the X-ray luminosity $L_{\rm X}$ (see also Figure\ \ref{fig:SEDth}), $\nu_{\rm shift}$ only increases by a factor of $\sim 2$. This is mostly because, $\nu_{\rm shift}$ depends primarily on BH mass $\mbh$ and secondarily on the ratio $\dot{m}_{\rm jet}/\dot{m}_{\rm MAD}$ (which, in our application, is regulated by M03 FP), while $L_{\rm X}$ has a steep positive dependence on $\dot{m}_{\rm MAD}$.

\begin{figure}
  \includegraphics[width=0.95\columnwidth]{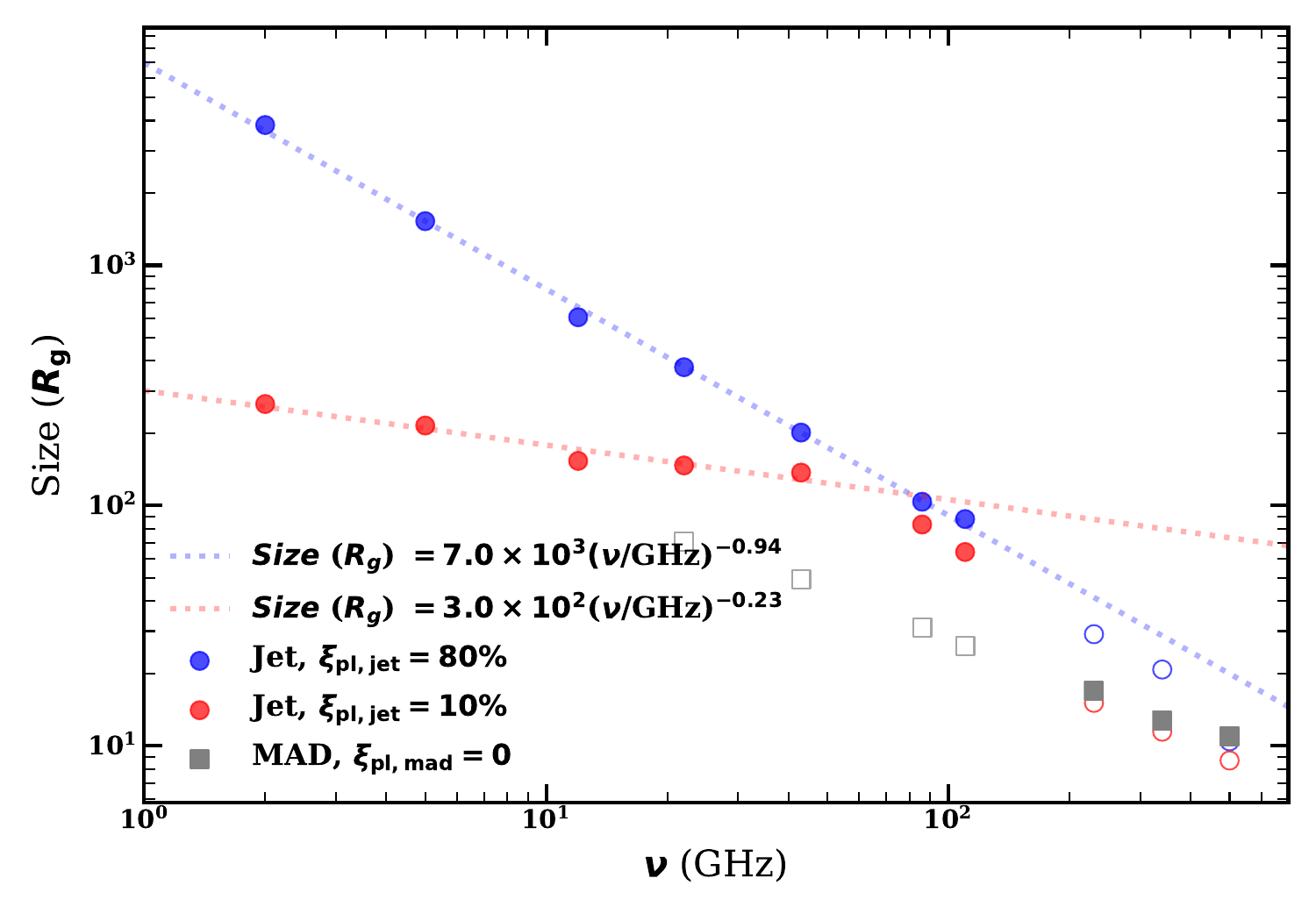}\\
  \includegraphics[width=0.95\columnwidth]{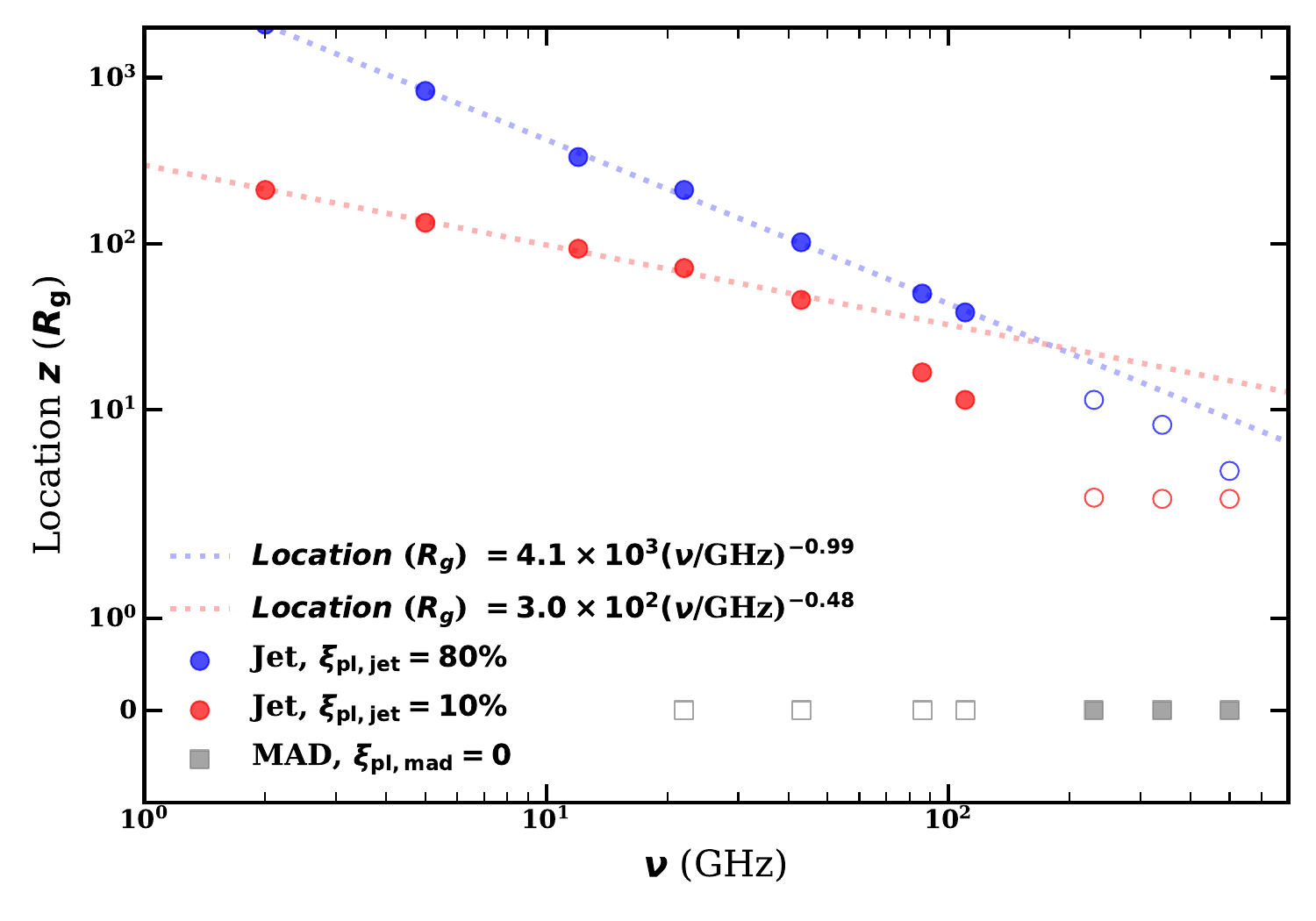}
  \caption{The radio core size ({\it top panel}) and location ({\it bottom panel}) of LLAGNs under the MAD.th-jet model.  We take $\dot{m}_0=1\times10^{-4}$ and a M03 FP, i.e. $\dot{m}_{\rm jet}=8.1\times10^{-7}$ for $\xi_{\rm pl, jet}=80\%$ (shown by blue circles) and $\dot{m}_{\rm jet}=1.03\times10^{-6}$ for $\xi_{\rm pl, jet}=10\%$ (shown by red circles). The predicted radio size and location of MAD is shown by gray squares. Filled and open properties of the symbols represent whether its emission dominates or not in the total spectrum at this frequency. For completeness, the dashed lines with the same color to the symbols show the power-law fits of the jet part.}
  \label{fig:size_mad.th_jet}
\end{figure}

We now focus on the location and size of the radio core emission. For this purpose, we take a system that the mass accretion rate is $\dot{m}_0=1\times10^{-4}$ (the X-ray luminosity $L_{\rm X} = 1.33\times 10^{40}\ergs$) and it also follows the M03 FP, i.e., the green curves in Figure \ref{fig:SEDth}. We then generate a series of images to calculate the location and size size of each component as a function of observed frequency.

Figure~\ref{fig:size_mad.th_jet} shows the resolved size (top panel) and the central location (bottom panel) of the radio emission at different frequencies. As shown in this plot, we consider two situations, one for $\xi_{\rm pl, jet}=10\%$ (most electrons are in thermal distribution, red circles) and the other for $\xi_{\rm pl, jet}=80\%$ (most electrons are in power-law distribution, blue circles). Filled (open) marks the jet (MAD) dominance in the total SED. Clearly, if most electrons in jet are in power-law (e.g., $\xi_{\rm pl, jet}=80\%$, blue circles), the jet size will follow $d_{\rm jet}\sim \nu^{-0.94}$, consistent with observations of LLAGN M104 \citep{Yan2024} and the prediction of BZ-jet \citep{Blandford1977, Blandford1979}. Actually, the jet size show no significant differences if $\xi_{\rm pl, jet}\ga 50\%$. On the other hand, if most electrons in jet are thermal, a much shallower profile is observed, i.e., $d_{\rm jet}\sim \nu^{-0.23}$ for the $\xi_{\rm pl,jet}=10\%$ case. In this work, our fiducial jet model takes $\xi_{\rm pl, jet} = 80\%$ (adopted as a representative of the $50-100\%$ range), to ensures a  $d_{\rm jet}\sim \nu^{-1}$ size-frequency relationship below $\sim 30$ GHz. Obviously, high spatial observations of other LLAGNs in the future may help constrain the value of $\xi_{\rm pl,jet}$, which may provide a hint on the particle acceleration mechanism in jet (\citealt{Yang2024}, and references therein).

In reality, the size and location of the radio core emission should be considered as a flux-weighted one. As shown in the bottom panel of Figure~\ref{fig:size_mad.th_jet}, due to the core-shift effect of jet, the location of peak jet emission moves inward quickly ($a \propto \nu^{-0.99} \sim \nu^{-1}$, see also \citealt{Blandford1979}). Consequently, at frequency $\nu_{\rm shift}$ where jet has the same flux to that of MAD, it locates very close to the central BH, i.e., $a_{\rm jet}\sim30 R_{\rm g}$ (equivalently, $\sim 0.03$mas for a $10^9\msun$ BH at a distance of 10 Mpc.) Considering the size of each component, this implies that technically we cannot resolve the two components (one for jet and the other for MAD) by current radio facilities. Separations of two radio components in faint LLAGNs (lower $\dot{M}$, thus lower $\nu_{\rm shift}$, see Fig. \ref{fig:nu_shift} or equivalently, larger $a$) in local Universe may still be resolvable. Even for LLAGNs that are moderately bright, it is still possible to argue the co-existence of relativistic jet and hot accretion flow (MAD here) if an elongated compact structure is observed.


\subsection{MAD.pl Case: Impact of Power-Law Electrons in MAD} \label{sec:size_LLAGN_MAD.pl}

One unsolved fundamental question in accretion theory is its energy dissipation, i.e., how is (part of) the released gravitational energy converted into thermal energy of gas, which eventually lights the environment through radiative process? Among all proposed mechanisms, magnetic reconnection and MHD turbulence are more promising (see \citealt{Xie2023} and references therein for a recent discussion). The energy dissipation plays an important role, not only in determining the fraction of energy that heats the electrons directly \citep{Sharma2007, Howes2010}, but also the efficiency in particle acceleration (that generates PL electrons, see e.g., \citealt{Sironi2011, Guo2014, Chael2018}). The magnetic reconnection process is extensively investigated through particle-in-cell simulations of the current-sheet regions, caused by reconnection (see e.g., \citealt{Ding2010}).

Impact of PL electrons in accretion systems has been extensively considered in Sgr A*, M87 and some other LLAGNs (e.g., \citealt{Yuan2003, Yuan2004, BandX2019, EHT2019, EHT2022}) and X-ray binaries (\citealt{Veledina2015} and references therein). 

As noted in \citet{Mahadevan1997}, the synchrotron self-absorption couples electrons efficiently at a fairly low accretion rate. Their estimation, based on conventional weak-magnetic-field hot accretion theory, is that the electrons remain thermal if $\dot{m} \ga 10^{-4}\alpha_{\rm vis}^2/\theta_{\rm e}$. This process is further emphasized in \citet{Malzac2009}, where it is renamed as a ``synchrotron boiler'' effect, i.e., synchrotron radiation ($\propto \gamma^2$) relaxes the electron distribution from PL towards a relativistic-Maxwellian thermal distribution with a weak high-energy tail.

\begin{figure}
  \includegraphics[width=0.95\columnwidth]{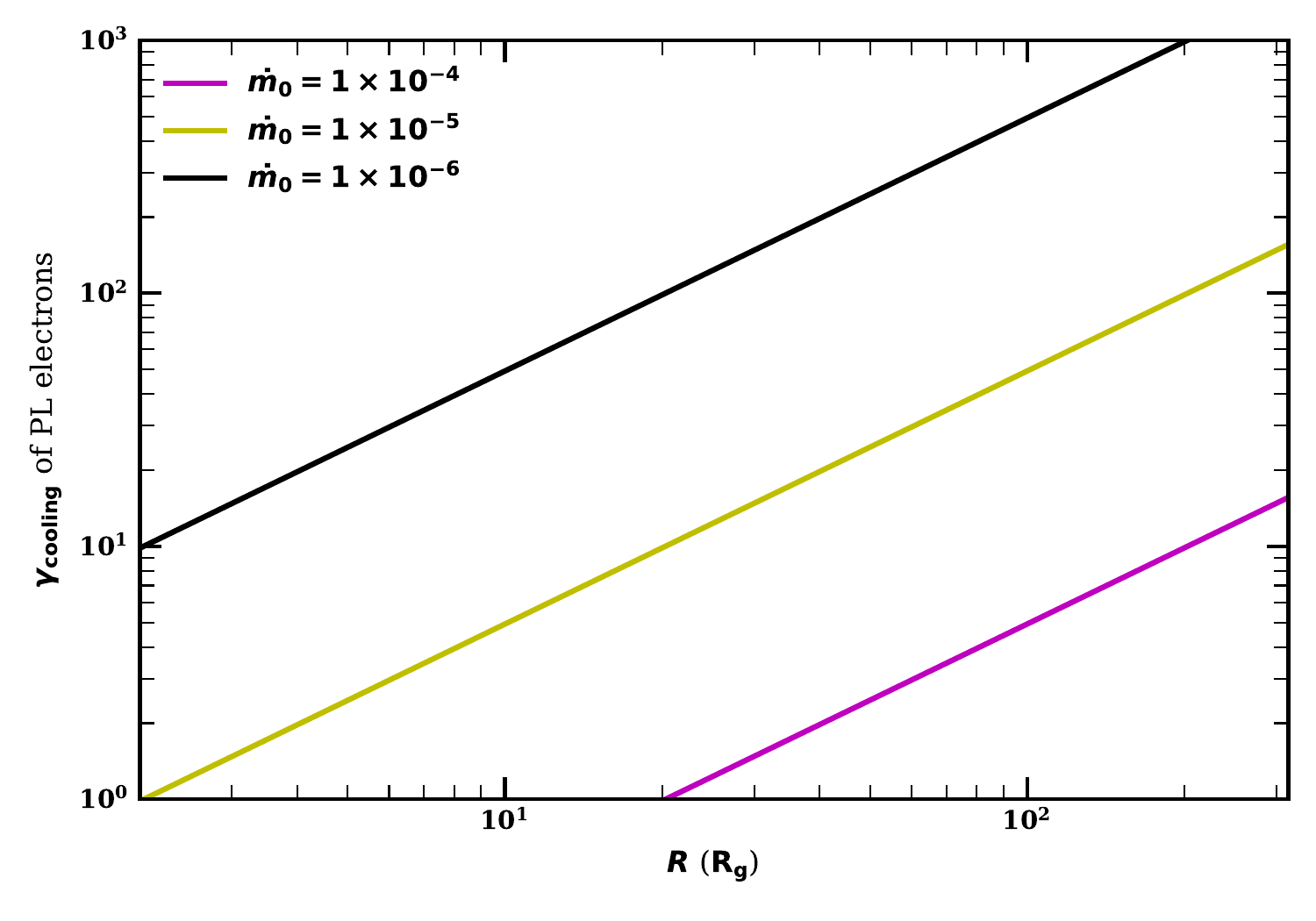}
  \caption{Estimated Lorentz factor due to synchrotron cooling effect ($\gamma_{\rm cooling}$) of PL electrons in MAD as a function of radius $R$. Lines from bottom to top represent MADs whose $\dot{m}_0=1\times10^{-4}$, $1\times10^{-5}$, and $1\times10^{-6}$, respectively. Clearly, `synchrotron boiler' effect cools most PL electrons, even at accretion rates as low as $\dot{m}_0\approx 1\times10^{-5}$ (the corresponding X-ray luminosity is $L_X\approx 1.44\times 10^{37}\ergs$).}
  \label{fig:gamma_cooling}
\end{figure}

Here we re-visit this problem in MAD picture. We first provide a crude estimation on $\gamma_{\rm cooling}$ under a simplified MAD model. The gas-to-magnetic plasma $\beta$-parameter is defined as $\beta_{\rm tot} = p_{\rm gas}/p_{\rm mag}$. The scale-height of MAD is expressed as $H =  c_s/\Omega_{\mathrm{K}}$ (where sound speed $c_s = \left(p_{\rm tot}/\rho\right)^{1/2}$), and the total (turbulent and global) magnetic field is $B_{\rm tot}^2/8\pi = p_{\rm tot}/(1+\beta_{\rm tot})$. The mass accretion rate and the radial velocity are, respectively,
\begin{eqnarray}
\dot{M} & \approx & -4\pi RH\rho V_{\rm R},\nonumber\\
V_{\rm R} & \approx & -\alpha_{\rm vis} (H/R)^2\, R\Omega_{\mathrm{K}}.\label{eq:dyn}
\end{eqnarray}
With the above formulae, we re-express the cooling Lorentz factor of MAD as (aspect ratio $H/R$ is kept explicitly), 
\begin{eqnarray}
    \gamma_{\rm cooling} & \approx & 3\pi\alpha_{\rm vis}^2 (1+\beta_{\rm tot}) \frac{m_{\rm e} c}{\sigma_{\rm T}} \left(\frac{H}{R}\right)^3 \frac{R}{\dot{M}}\nonumber\\
    & \propto & \alpha_{\rm vis}^2(1+\beta_{\rm tot}) \left(\frac{H}{R}\right)^3 \frac{R/R_{\rm g}}{\dot{M}/\dot{M}_{\rm Edd}}. \label{eq:gamma_cooling_mdot}
\end{eqnarray}
From Equation (\ref{eq:gamma_cooling_mdot}), we find that $\gamma_{\rm cooling}$ is independent of BH mass, i.e., AGNs and BH binaries share the same $\gamma_{\rm cooling}$. Besides, a positive (negative) relation with radius (accretion rate) can also be observed. We caution that, the above expression only considers (local) synchrotron cooling effect, inverse Compton scattering to those local self-generated photons and those global propagated pass-by photons is not taken into account (e.g., \citealt{Xie2010}). Besides, most of the energy is released close to BH, possible existence of the high-energy electrons at large radius should not produce much radiation. In this work, we limit ourselves to $R < 200 R_{\rm g}$. 

Numerical calculations of $\gamma_{\rm cooling}$ based on Equation (\ref{eq:gamma_cooling_mdot}) are shown in Figure \ref{fig:gamma_cooling}. In our calculations, we adopt $\alpha_{\rm vis}=0.3$, $\beta_{\rm tot} = 6$, and fix the aspect ratio of MAD $H/R=0.5$ (equivalently, we neglect the dependence of $H/R$ on $\dot{M}$). Three accretion rates are considered, i.e., lines from bottom to top represent $\dot{m}_0 = 1\times10^{-4}$, $1\times10^{-5}$, and $1\times10^{-6}$. As shown by this plot, most of the PL electrons will be cooled down by the `synchrotron boiler' effect, even at a fairly low accretion rate (i.e., $\dot{m}_0\approx (1-5)\times10^{-5}$; the corresponding X-ray luminosity is $L_X\approx (1.4-10)\times 10^{37} \ergs$). In practice, when radiative cooling is considered, the minimum energy of PL electrons is linked to the energy of thermal electrons, as calculated from $d(\gamma n_{\rm th}(\gamma))/d\gamma = 0$. We assume that, through the `synchrotron boiler' effect, electrons whose energy is below this value are totally `thermalized'.


\begin{figure}
  \centering
  \includegraphics[width=0.955\columnwidth]{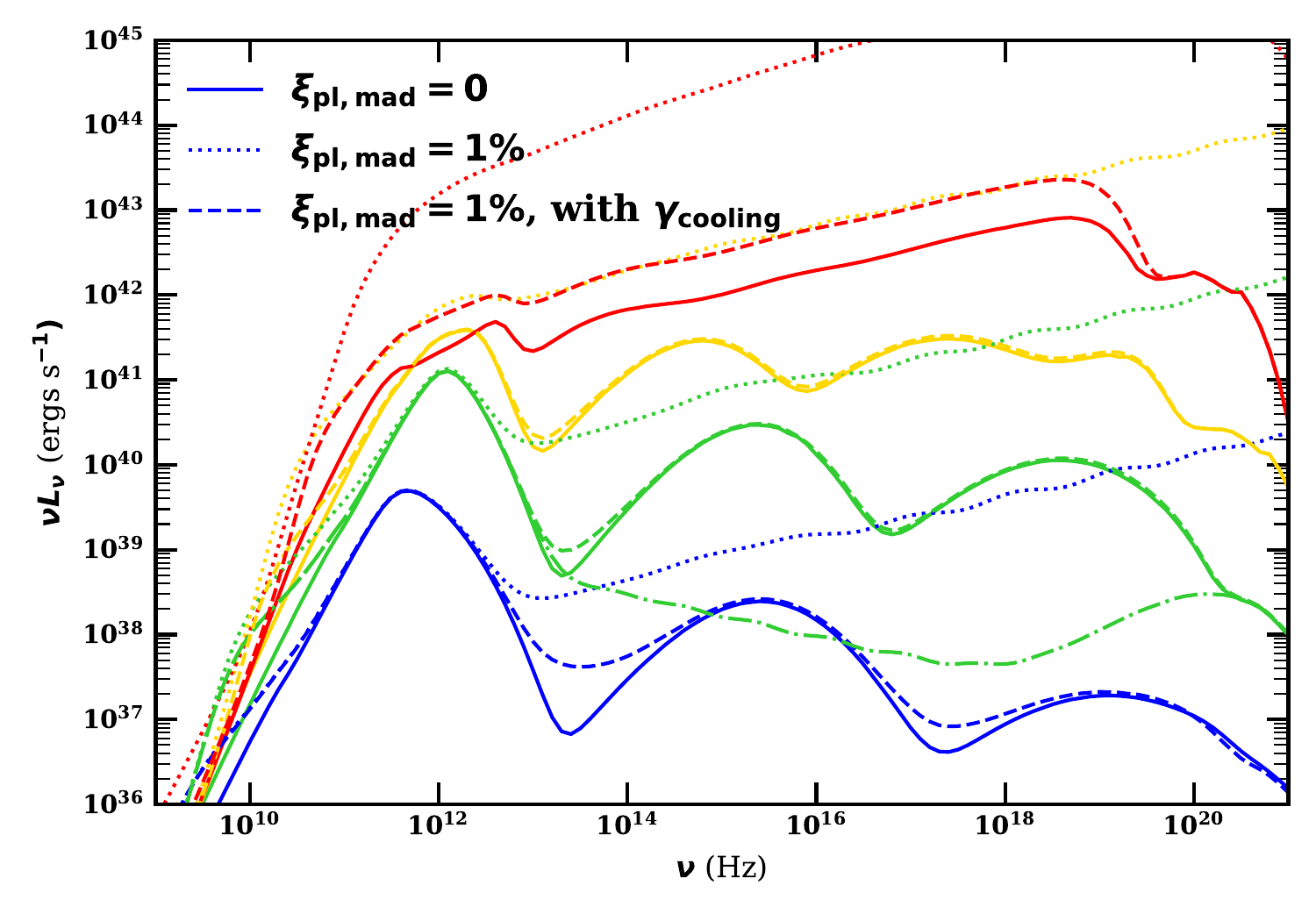}
 \caption{Broadband SED of MAD models under different PL electron setups. From top to bottom, lines with the same color represent MADs whose accretion rates are $\dot{m}_0=1\times10^{-2}$, $1\times10^{-3}$,  $1\times10^{-4}$, and $1\times10^{-5}$, respectively. For each accretion rate, three PL electron setups are considered (lines with the same color), i.e., solid for the $\xi_{\rm pl, mad} = 0$ (MAD.th case), dotted for the fixed distribution PL case ($\xi_{\rm pl, mad} = 1\%$, radiative cooling to PL electrons excluded, a $\nu L_\nu \propto \nu^{\sim +0.23}$ spectrum is observed at $\nu\gtrsim\times10^{14}$ Hz), and dashed for the self-consistent PL case ($\xi_{\rm pl, mad} = 1\%$, and radiative cooling to PL electrons included). For comparative purpose, the green dot-dashed curve shows the spectrum without inverse Compton, for the case of PL electrons with $\gamma_{\rm cooling}$. A $\nu L_\nu \propto \nu^{\sim -0.246}$ spectrum is indeed observed at $10^{14}$-$10^{17}$ Hz. Above $\sim 10^{18}$ Hz is the bremsstrahlung bump.}
  \label{fig:SED_pl}
\end{figure}

\begin{figure}
\includegraphics[width=0.95\columnwidth]
  {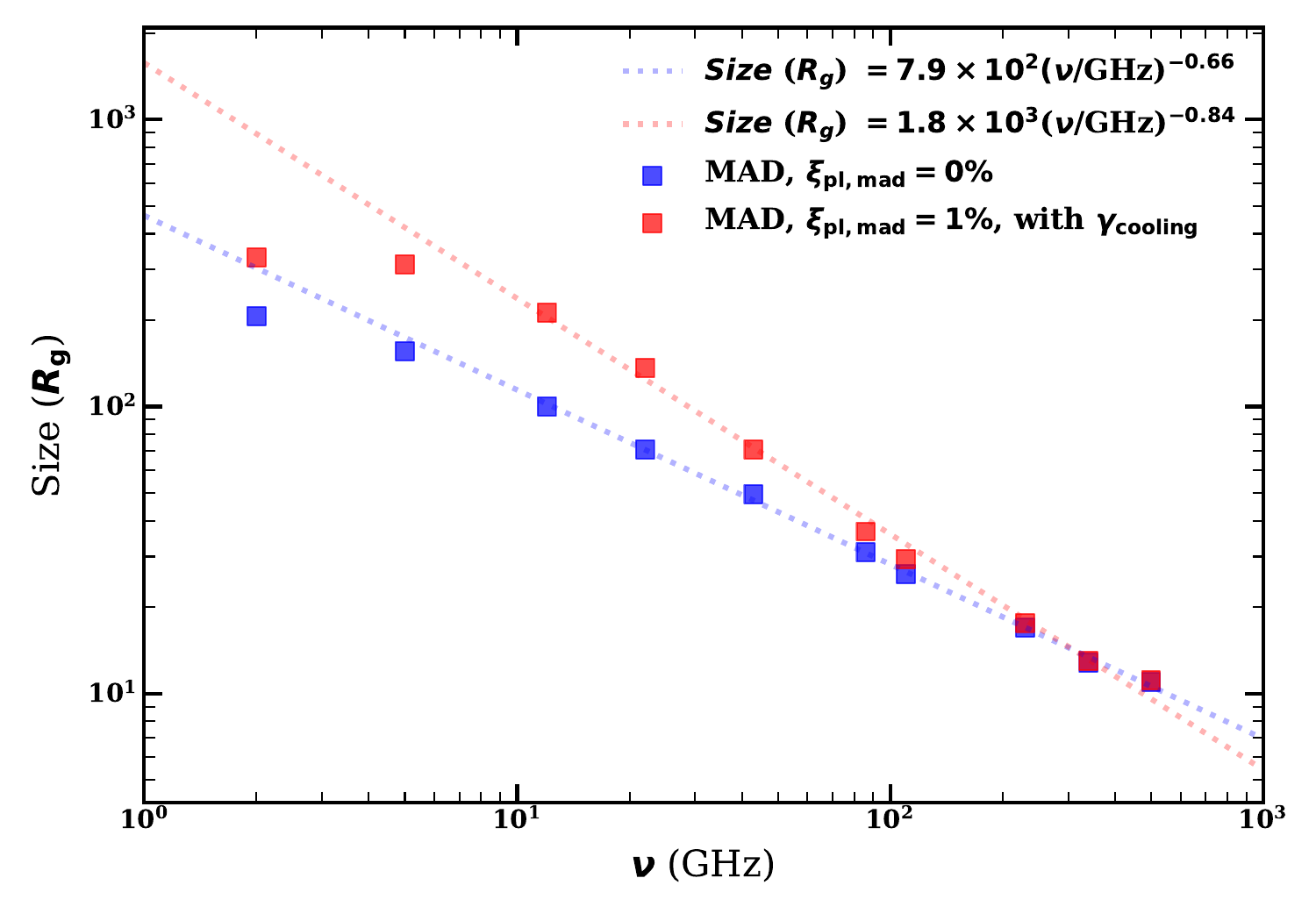}
  \caption{Radio size of MAD in LLAGNs. Here we take cases of $\dot{m}_0 = 1\times10^{-4}$ (see light-green curves in the left panel for the SED) for illustration. Blue and red filled squares represent the radio size of MAD with $\xi_{\rm pl, mad}=0$ and $\xi_{\rm pl, mad} = 1\%$ (cooling included), respectively. We emphasize that the radio size of LLAGNs at low frequency should be dominated by the jet component (if exist).}
  \label{fig:SED_size_pl}
\end{figure}

Figure~\ref{fig:SED_pl} shows the SED of the MAD based on three electron population scenarios: (i) thermal electrons only ($\xi_{\rm pl, mad} = 0$, shown by solid curves. MAD.th case. See also solid curves in Figure \ref{fig:SEDth}); (ii) PL electrons intrinsically received a fraction of $\xi_{\rm pl, mad}=1\%$ thermal energy. but radiative cooling to these PL electrons excluded (dotted curves); and (iii) $\xi_{\rm pl, mad}=1\%$, and radiative cooling to PL electrons is considered self-consistently (dashed curves; self-consistent MAD.pl case hereafter). As demonstrated in Fig.~\ref{fig:gamma_cooling}, physically the case (ii) is equivalent to that, the accelerator like magnetic reconnection is so-efficient that it can immediately complement the loss of high-$\gamma$ power-law electrons, i.e., the real total energy received by those PL electrons are unrealistically increased.

We first focus on low-luminosity cases of $\dot{m}_0 = 1\times10^{-5}$ and $1\times 10^{-4}$, where the scattering optical depth is low (at a given radius, electron scattering optical depth $\tau_{\rm es}\approx 2\rho H \propto \alpha_{\rm vis}^{-1} (H/R)^{-2} \dot{M}$, see Equation (\ref{eq:dyn}).) thus high-order inverse Compton scattering is not very important. As shown by the dotted curve in this plot, we can clearly find that, through synchrotron, the existence of PL electrons in MAD will highly enhance the emission at low-frequency radio (see also \citealt{Yuan2003} for Sgr A*)\footnote{Technically, additional low-frequency radio emission from relativistic jet should be included, to meet the MAD-jet coupling setup, where we take the M03 FP.} and most significantly at the broad infrared-X-ray band. Clearly, synchrotron emission of the PL electrons without cooling should follow $\nu L_\nu \propto \nu^{(3-p_{\rm pl,mad})/2}\approx \nu^{0.25}$ \citep{Rybicki1979}. This is clearly shown in the figure. In other words, without considering radiative cooling to PL electrons, we will observe a hard X-ray spectrum with a photon index $\Gamma = (p_{\rm pl,mad}+1)/2 \approx 1.75$. Due to low number density at the high-$\gamma$ end,
their contribution in the inverse Compton scattering process is negligible in the observed wavebands from radio up to X-rays \citep{Rybicki1979}. Finally we caution that, without radiative cooling, our fixed fraction of PL electrons $\xi_{\rm pl, mad}$ may be too large, that $L_{\rm bol}/\dot{M} c^2 > 100\%$ (depending on $\gamma_{\rm pl, max}$). In reality only a much lower value of $\xi_{\rm pl, mad}$ may be physically plausible.

However, once radiative cooling is included, most of the high-energy electrons will be cooled down effectively (see Figure \ref{fig:gamma_cooling}), and their contribution to total SED turns out to be much weaker, as shown by comparing dashed curves to dotted ones. This suggests that considering radiative cooling to PL electrons is of crucial importance in order to develop a physically-reasonable model, as long as $\dot{m}_0 > 1\times10^{-7}$ (see Equation (\ref{eq:gamma_cooling_mdot}), and \citealt{Mahadevan1997}). For comparative purpose, we also show in Figure \ref{fig:SED_pl} by the green dot-dashed curve the SED without Compton scattering, for the case of PL electrons with $\gamma_{\rm cooling}$. We indeed observed a $\nu L_\nu \propto \nu^{\sim -0.23}$ SED between $10^{14}$ Hz and $10^{17}$ Hz, consistent with theoretical expectation of synchrotron emission by PL electrons that follow a $\gamma^{-(p_{\rm pl, mad}+1)} = \gamma^{-3.5}$ distribution.

We now compare the MAD.th-jet model (see Fig.~\ref{fig:SEDth}) to the self-consistent MAD.pl model. One interesting result is that, although in both cases the PL electrons are responsible for the low-frequency radio emission, the spectrum of MAD.pl itself is much steeper than the jet component. This directly reflects the dynamical differences between MAD and jet, which are quite obvious. Although challenging, a detailed modeling of observational data from a broadband simultaneous monitoring in the whole radio band, from $\sim 1$ GHz up to $\sim 40-100$ GHz, may help to find signatures of PL electrons within MAD itself. 

The dashed curves in Figure \ref{fig:mad_alpha_s} show the spectral index $\alpha_{\rm s}$ of the self-consistent MAD.pl models. Compared to the MAD.th case, PL electrons will enhance the low-frequency radio emission. Consequently, $\alpha_{\rm s}$ measured in the 20-500 GHz band will be smaller in the MAD.pl case compared to the MAD.th case (blue dashed vs. blue solid). At extremely high accretion rates, most of the PL electrons will be cooled down (see Figure \ref{fig:gamma_cooling}). Consequently, it will mostly recover the MAD.th case (without PL electrons). This is clearly observed at the right-end of this plot, where $\alpha_{\rm s}$ of MAD.th agrees with that of self-consistent MAD.pl.

Figure~\ref{fig:SED_size_pl} shows the radio size of MAD (in units of $R_{\rm g}$) in LLAGNs. Here we take cases of $\dot{m}_0 = 1\times10^{-4}$ (see light-green curves in the left panel for the SED) for illustration. Blue and red filled squares represent the radio size of MAD.th and self-consistent MAD.pl cases with $\xi_{\rm pl, mad} = 1\%$, respectively. Note that this plot only considers emission from MAD. If relativistic jet exists, the radio band below $\nu_{\rm shift}$ (see Figure \ref{fig:nu_shift}) will be primarily from jet. A simple power-law fit to the size-frequency relationship between $10$ GHz and $500$ GHz of MAD emission reads, respectively, 
\begin{eqnarray}
d_{\rm mad}(\nu) & \approx & 7.9\times 10^2R_{\rm g}\,(\nu/\,{\rm GHz})^{-0.66}, \hspace{0.2cm} (\xi_{\rm pl, mad} = 0),\\
d_{\rm mad}(\nu) & \approx & 1.8\times10^3 R_{\rm g}\,(\nu/\,{\rm GHz})^{-0.84}, \hspace{0.2cm} (\xi_{\rm pl, mad} = 1\%).
\end{eqnarray}
As shown in this plot, in the 10-100 GHz regime, the size of two cases are quite similar. The MAD.pl case has a shallower slope, although the difference is insignificant. This reflects the fact that PL electrons, if exist, mostly contribute to radio emission at low-frequencies.

\begin{figure}
\centering
\includegraphics[width=0.95\columnwidth]{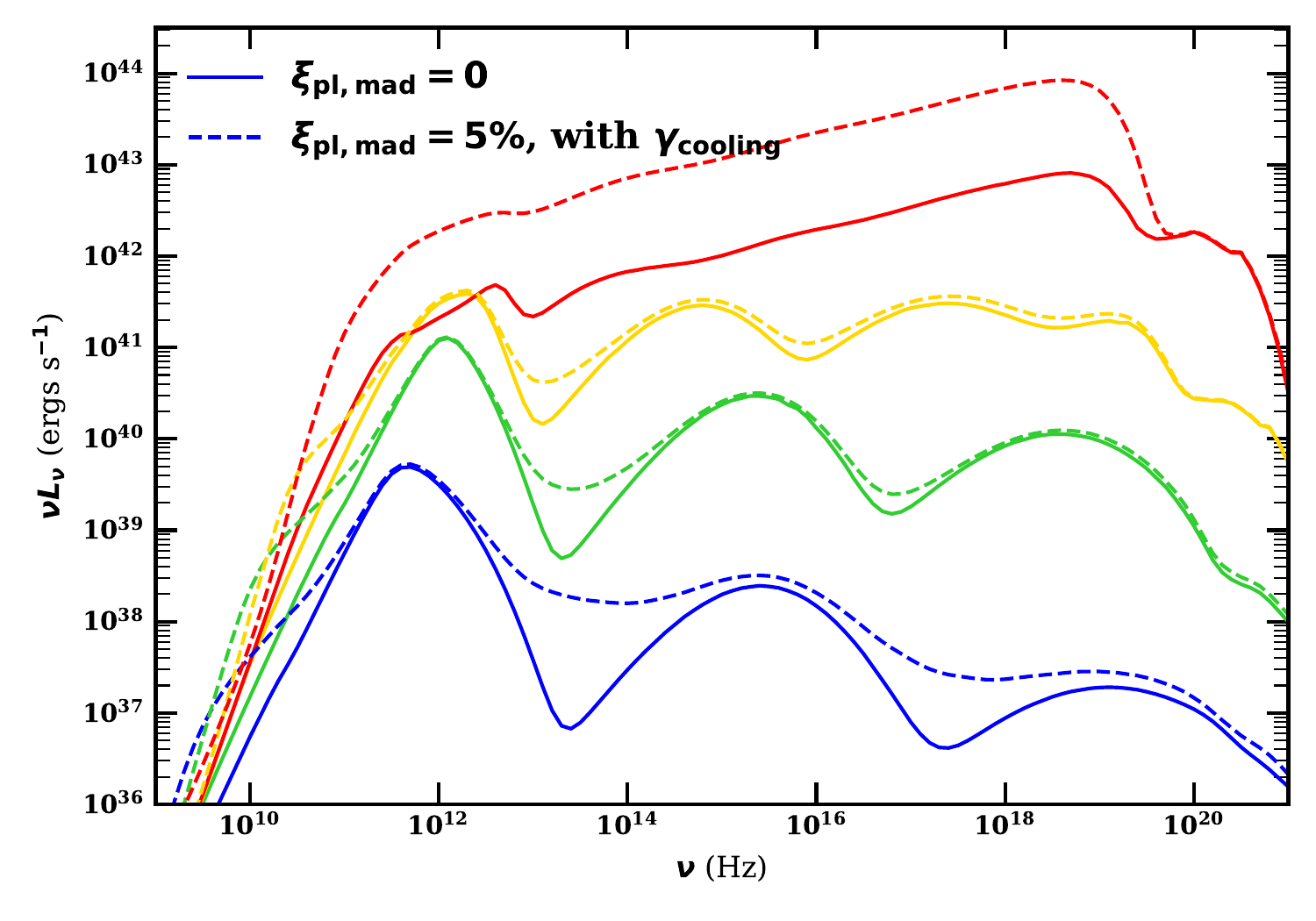}
  \caption{Broadband SED of MAD models under different setups of PL electrons. From top to bottom, lines with the same color represent MADs whose accretion rates are $\dot{m}_0=1\times10^{-2}$, $1\times10^{-3}$,  $1\times10^{-4}$, and $1\times10^{-5}$, respectively. Solid and dashed lines are for cases that only include thermal electrons ($\xi_{\rm pl, mad} = 0$), and self-consistently include $\xi_{\rm pl, mad} = 5\%$ PL electrons, respectively.}
  \label{fig:SEDpl_xi5}
\end{figure}

Considering the uncertainty in $\xi_{\rm pl, mad}$, below we investigate its impact on a MAD.pl case with a higher value of $\xi_{\rm pl, mad} = 5\%$. The spectrum is shown in Fig.~\ref{fig:SEDpl_xi5}. Four accretion rates are considered, i.e.,  $\dot{m}_0 = 1\times10^{-5}, 1\times10^{-4}, 1\times10^{-3}$, and $1\times10^{-2}$, represented by dashed curves from bottom to top. For comparison, cases with $\xi_{\rm pl, mad} = 0$ (MAD.th) are also included, as shown by solid curves with the same color. From this plot, we find that the low-frequency radio bump due to PL electrons in MAD is much brighter than its corresponding cases with $\xi_{\rm pl, mad} =1\%$. Their impact on X-rays is also evident, especially at high accretion rates where the optical depths are systematically higher.

\section{summary}\label{sec:summary}

After decades of investigation, through synergies of various approaches that include broad-band spectrum, emission and absorption lines, variability and time lag analysis, the basic picture of the AGN structure has been developed. By increasing the baseline (e.g., VLBA, EVN) and extending to high-frequency bands (e.g., EHT), the spatial resolution in radio is highly improved. These high-resolution radio observations have significantly advanced our understanding of LLAGNs and black hole physics (e.g., \citealt{Bower2015, BandX2019, EHT2019, EHT2022, Yan2024, Algaba2024}). 

Based on a theoretical MAD-jet coupling model, this study systematically explores the spectrum and more importantly, the size and location of LLAGNs measured in radio band, with the aim of distinguishing the origin of radio emission at different frequencies. Our main findings can be summarized as follows:

\begin{itemize}

\item We reproduce the broadband spectrum and the frequency-dependent radio core size of LLAGN M104 based on the MAD-jet model, an update to \citet{Yan2024}.

\item For LLAGNs in the radio band, as frequency increases, we should observe a shift of the radio emission origin from jet to MAD. For a $\mbh = 10^9\msun$ LLAGN that follows the M03 FP, even though the 2-10 keV X-ray luminosity of LLAGN varies by a factor of $\sim$ $10^3$, $\nu_{\rm shift}$ varies by only a factor of $\sim$2 (see Figure \ref{fig:nu_shift}). Theoretically, the shift frequency $\nu_{\rm shift}$ depends primarily on BH mass $\mbh$, and secondarily on the ratio $\dot{m}_{\rm jet}/\dot{m}_{\rm MAD}$, i.e. reflects the competition between accretion and ejection.

\item The size and location of the jet component show a clear dependence on the energy fraction of PL electrons ($\xi_{\rm pl,jet}$), if we take a hadronic jet model and a large bulk velocity ($\Gamma_{\rm jet} \gtrsim 5$). $\xi_{\rm pl, jet}$ can be constrained in future by a detailed observations of the radio core size of LLAGNs at frequencies below $\nu_{\rm shift}$ ($\sim 100-250$ GHz for LLAGNs whose $\mbh = 10^9 \msun$).

\item For most LLAGNs and X-ray binaries in hard states that have $L_{\rm bol}/L_{\rm Edd} \gtrsim (3-8)\times10^{-6}$, PL electrons in MAD are predicted to be fairly weak, and MAD.th is most favored. We re-emphasize that, even though a significant fraction of PL electrons can be accelerated in-situ through magnetic reconnection, they will be cooled down immediately due to strong radiative cooling (so-called `synchrotron boiler' effect, see \citealt{Malzac2009}). A crude estimation of cooling Lorentz factor that only considers local self-synchrotron reads $\gamma_{\rm cooling} \propto (H/R)^{3}\, (R/R_{\rm g})\,(\dot{M}/\dot{M}_{\rm Edd})^{-1}$, independent of $\mbh$. This implies that including PL electrons without a self-consistent treatment on radiative cooling (equivalent to an immediate ``accelerator'' that implement high-$\gamma$ PL electrons) will significantly overestimate the outcome radiation, e.g., by a factor of $\sim 10$-100 even for $\xi_{\rm pl,mad}=1\%$.

\item Impact of PL electrons in MAD is investigated. Based on a self-consistent MAD.pl model with $\xi_{\rm pl, mad} = 1\%$, we find that the low-frequency radio and infrared becomes much brighter. However, due to strong radiative cooling, the whole system lacks sufficient high-energy electrons (see Figure~\ref{fig:gamma_cooling}). Consequently, emission at X-rays is only weakly enhanced. Besides, additional synchrotron emission by those PL electrons also changes (enlarges) slightly the radio core size of MAD. Considering the co-existence of jet emission, this is difficult to constrain.

\end{itemize}

\section*{Acknowledgements}
We appreciate the referee for a careful reading, and informative and insightful suggestions, which improve our model setup. This work was supported in part by the National Natural Science Foundation of China (NSFC Nos. 12373017, 12192220, and 12192223).



\bibliography{LLAGN_size.bib}{}
\bibliographystyle{aasjournalv7}




\end{CJK*}
\end{document}